\documentclass[nofootinbib,pra,twocolumn,showpacs,superscriptaddress,notitlepage,amsmath,amssymb]{revtex4-2}
\usepackage{amsmath,amssymb}
\usepackage{graphicx}
\usepackage[colorlinks=true,linkcolor=blue,citecolor=blue,urlcolor=blue]{hyperref}
\usepackage{xcolor}
\usepackage{pifont}
\usepackage{braket}
\usepackage{bm}

\usepackage[normalem]{ulem}

\usepackage{framed}
\definecolor{shadecolor}{RGB}{224,224,224}

\newcommand{\R}[1]{{\color{red} #1}}
\definecolor{DarkGreen}{RGB}{0,128,0}
\newcommand{\G}[1]{{\color{DarkGreen} #1}}
\newcommand{\B}[1]{{\color{blue} #1}}
\newcommand{\GG}{{\color{DarkGreen} G}}
\newcommand{\BB}{{\color{blue} B}}
\newcommand{\RR}{{\color{red} R }}

\newcommand{\trivalent}{
  \mathbin{
    \begin{tikzpicture}[scale=0.24,baseline=-0.1ex]
      \draw (0,0)--(0,1);
      \draw (0,0)--(-0.866,-0.5);
      \draw (0,0)--( 0.866,-0.5);
    \end{tikzpicture}
  }
}
\usepackage{tikz}
\newcommand{\topstring}{
  \mathbin{
    \begin{tikzpicture}[scale=0.24,baseline=-0.1ex]
      \draw (0,0)--(0,1);
    \end{tikzpicture}
  }
}
\newcommand{\bottomright}{
  \mathbin{
    \begin{tikzpicture}[scale=0.24,baseline=-0.1ex]
      \draw (0,0)--( 0.866,-0.5);
    \end{tikzpicture}
  }
}
\newcommand{\bottomleft}{
  \mathbin{
    \begin{tikzpicture}[scale=0.24,baseline=-0.1ex]
      \draw (0,0)--(-0.866,-0.5);
    \end{tikzpicture}
  }
}
\newcommand{\topright}{
  \mathbin{
    \begin{tikzpicture}[scale=0.24,baseline=-0.1ex]
      \draw (0,0)--(0,1);
      \draw (0,0)--( 0.866,-0.5);
    \end{tikzpicture}
  }
}
\newcommand{\topleft}{
  \mathbin{
    \begin{tikzpicture}[scale=0.24,baseline=-0.1ex]
      \draw (0,0)--(0,1);
      \draw (0,0)--(-0.866,-0.5);
    \end{tikzpicture}
  }
}
\usepackage{tikz}

\newcommand{\bivalent}{
  \mathbin{
    \begin{tikzpicture}[scale=0.24,baseline=-0.1ex]
      \draw (0,0.6)--(-0.5,-0.2);
      \draw (0,0.6)--( 0.5,-0.2);
    \end{tikzpicture}
  }
}

\usepackage{pgfplots}
\usetikzlibrary{fadings}
\usetikzlibrary{plotmarks}

\usetikzlibrary{shapes.geometric}

\pgfdeclareplotmark{redstar}{
\node[star,star point ratio=2.25,minimum size=0.4cm,inner sep=0,draw=white,solid,fill=orange] {};
}
\pgfdeclareplotmark{blackstar}{
\node[star,star point ratio=2.25,minimum size=0.4cm,inner sep=0,draw=white,solid,fill=black] {};
}

\pgfdeclareplotmark{myblack}{
\node[square/.style={regular polygon,regular polygon sides=4},minimum size=0.18cm,inner sep=0,draw=white,solid,fill=black] {};
}
\pgfdeclareplotmark{myorange}{
\draw[thick,fill=yellow] (0,0) -- (30:0.12cm) arc (30:330:0.12cm) -- cycle;
}

\tikzset{>=stealth}

\begin{document}

\title{Intrinsic Heralding and Optimal Decoders for Non-Abelian Topological Order}

\author{Dian Jing}
\email{rossoneri@uchicago.edu}
\affiliation{Department of Physics, University of Chicago, Chicago, Illinois 60637, USA}
\affiliation{Pritzker School of Molecular Engineering, University of Chicago, Chicago, Illinois 60637, USA}

\author{Pablo Sala}
\affiliation{Department of Physics and Institute for Quantum Information and Matter, California Institute of Technology,
Pasadena, California 91125, USA}
\affiliation{Walter Burke Institute for Theoretical Physics, California Institute of Technology, Pasadena, California 91125, USA}

\author{Liang Jiang}
\affiliation{Pritzker School of Molecular Engineering, University of Chicago, Chicago, Illinois 60637, USA}

\author{Ruben Verresen}
\email{verresen@uchicago.edu}
\affiliation{Pritzker School of Molecular Engineering, University of Chicago, Chicago, Illinois 60637, USA}

\date{\today}

\begin{abstract}
Topological order (TO) provides a natural platform for storing and manipulating quantum information. However, its stability to noise has only been systematically understood for Abelian TOs. In this Letter, we exploit the nondeterministic fusion of non-Abelian anyons to inform active error-correction and design decoders where the fusion products, instead of flag qubits, herald the noise. This intrinsic heralding offers improved thresholds over those of Abelian counterparts. Furthermore, we use Bayesian inference to obtain a statistical mechanics model for fixed-point non-Abelian TOs with perfect measurements under any noise model, which yields the optimal threshold conditioned on measuring anyon syndromes.
We numerically illustrate these results for $D_4 \cong \mathbb Z_4 \rtimes \mathbb Z_2$ TO. In particular, for non-Abelian charge noise and perfect syndrome measurement, we find a conditioned optimal threshold $p_c = 0.218(1)$, whereas an intrinsically heralded minimum-weight perfect matching (MWPM) decoder already gives $p_c = 0.20842(2)$, outperforming standard honeycomb-lattice MWPM with $p_c = 0.15860(1)$. Our Letter highlights how non-Abelian properties can enhance stability, rather than reduce it, and discusses potential generalizations for achieving fault tolerance.
\end{abstract}

\maketitle

\begin{figure}[t]
\begin{tikzpicture}
\node at (0,0) {
\includegraphics[scale=0.7]{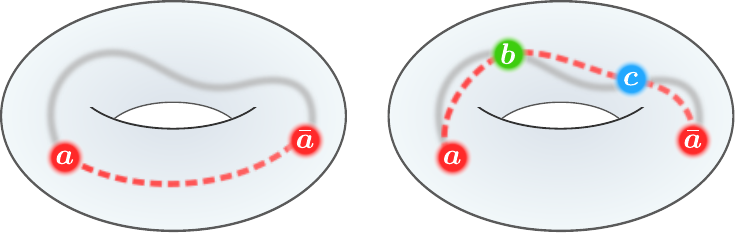}
};
\node at (-2.3,1.7) {without heralding};
\node at (2.3,1.7) {intrinsic heralding};
\node[color=gray] at (-1.6,0.65) {error};
\node[color=red] at (-2.3,-0.5) {best guess};
\node at (0,-1.6) {$a \times \Bar{a} = 1+b+c+\cdots$};
\end{tikzpicture}
\caption{\textbf{Intrinsic heralding from non-Abelian anyons.} A constant-depth error string not only creates a pair of non-Abelian anyons at its endpoints, but also leaves behind a superposition over possible fusion outcomes along its path. Intermediate anyon syndromes can be extracted by collapsing this superposition, providing information about the original error path and thereby improving error correction. Such intrinsic heralding arises from the nondeterministic fusion of non-Abelian anyons, without the need for flag qubits.}
\label{heralding}
\end{figure}

\begin{figure}[t]
\begin{tikzpicture}
\node at (1.9,1.1) {using $[2]$ syndromes};
\node at (6,1.1) {using all syndromes};
\node at (1,0){\includegraphics[scale=0.4]{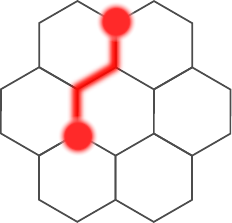}};
\node at (2.8,0){\includegraphics[scale=0.4]{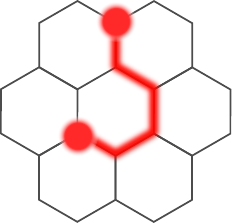}};
\node at (5.2,0){\includegraphics[scale=0.4]{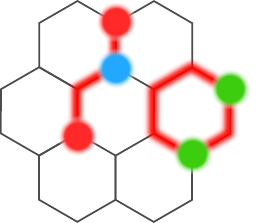}};
\node at (7,0){\includegraphics[scale=0.4]{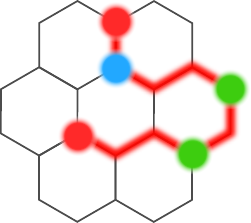}};
\node at (0,-1) {$P(E) \propto$};
\node at (1.1,-1) {$t^{3}$};
\node at (2.8,-1) {$t^{5}$};
\node at (5.2,-1) {$t^{3+6}$};
\node at (7,-1) {$t^{9}$};
\node[orange] at (3.9,-1.9) {$ P(\bm s|E) =$};
\node at (5.2-0.1,-1.4) {\footnotesize $\times$};
\node[orange] at (5.2-0.1,-1.9) {$\frac{2^2}{2^{2+6}}$};
\node at (7-0.1,-1.4) {\footnotesize $\times$};
\node[orange] at (7-0.1,-1.9) {$\frac{2^0}{2^{8}}$};
\draw[->,opacity=0.2,line width=1,path fading=south] (2.3,-3.2) to[out=120,in=-90] (2,-1.5);
\draw[->,orange,opacity=0.4,line width=1,path fading=south] (6,-3.5) to[out=70,in=-90] (6.2,-2.5);
\node at (3.6,-4){
\begin{tikzpicture}
\begin{axis}[axis x line=middle,axis y line=middle,xmin=0,xmax=0.24,ymin=0,ymax=4.2,height=4.5cm,width=8cm,xlabel={$p$},ylabel={$T$},label style={font=\normalsize},x label style={at={(axis description cs:1.08,0)},anchor=east},y label style={at={(axis description cs:0,1.15)},anchor=north},axis line style={->},label style={font=\normalsize},tick label style={font=\footnotesize,/pgf/number format/.cd,fixed,precision=3},xtick={0.05,0.1,0.159,0.208},ytick={1,2,3,3.64}]
\addplot[black,dashed,domain=0.0001:0.9,samples=200]({x},{2/ln(1/x-1)});
\addplot[smooth,thick,samples=50,tension=1] coordinates {( 0,3.64) ( 0.1033,2.498) ( 0.1642,1.229)}; 

\addplot[samples=50,thick] coordinates {( 0.1642,1.229) ( 0.1586,0)};
\addplot[mark=*,mark size=0.03cm] coordinates {( 0,3.64)};
\addplot[mark=*,mark size=0.03cm] coordinates {( 0.1033,2.498)};
\addplot[mark=blackstar,orange,mark size=0.12cm] coordinates {( 0.1642,1.229)};
\addplot[mark=myblack] coordinates {( 0.1586,0)};

\addplot[smooth,orange,thick,samples=50,tension=1,path fading=north] coordinates {(0.145,7.65) (0.186,3.82) (0.2177,1.5636)}; 

\addplot[samples=50,orange,thick] coordinates {(0.2177,1.5636) (0.2120,0)};

\addplot[mark=*,orange,mark size=0.03cm] coordinates {(0.186,3.82)};
\addplot[mark=redstar,red] coordinates {(0.2177,1.5636)};

\addplot[mark=myorange] coordinates {(0.2084,0)};

\addplot[mark=*,orange,mark size=0.03cm] coordinates {(0.2120,0)};
\end{axis}
\end{tikzpicture}
};
\end{tikzpicture}
\caption{\textbf{Error-correction thresholds for the TO $D_4$ under non-Abelian charge noise.} The phase diagram of the stat-mech models to which the error-correction problems of the $D_4$ TO under non-Abelian charge noise are mapped, with representative snapshots where anyons occupy the vertices and error strings trace the edges of the honeycomb lattice. The black phase boundary corresponds to the random-bond Ising model associated with unheralded matching decoders that only see the non-Abelian charge anyon (red dot in snapshots) and considers only $P(E) \propto t^{|E|}$ with $t = e^{-2\beta}$; the black square and star mark the thresholds of the MWPM and maximum-likelihood decoders (Eq.~\eqref{TC}), respectively. The yellow symbol is the intrinsically heralded MWPM decoder, where the error string is forced to pass through all Abelian charge fusion products (blue and green dots in snapshots), with threshold $p_c = 0.20842(2)$. The orange phase boundary corresponds to the threshold using the full set of syndromes, where each snapshot is weighted by $P(\bm s|E) P(E)$ (Eq.~\eqref{Bayes}). Here, $P(\bm s |E)$ (Eq.~\eqref{Probability}) includes a factor of $\frac{1}{2}$ for each vertex of the error string not passing through a non-Abelian anyon, and a factor of $2$ for each Abelian parity constraint, as illustrated by the last two snapshots. The optimal decoder is marked by the orange star along the Nishimori line (dashed), $t = \frac{p}{1-p}$, with $p_c = 0.218(1)$. Details on stat-mech models, calculation of $P(\bm{s}|E)$, and numerical simulations can be found in Ref.~\onlinecite{SM}.}
\label{phase_diagram}
\end{figure}

\indent \textit{Introduction.} Topological orders (TOs) are long-range entangled topological phases characterized by ground state degeneracy and anyon excitations \cite{Leinaas1977, Goldin81, Wilczek82, Goldin85, Moore1989, Wen90, Wen91, MOORE1991362, Einarsson95}. They have been exploited to encode and manipulate quantum information thanks to their robustness against local noise \cite{KITAEV20032, Freedman2002, Freedman06, RevModPhys2008, Dennis, RMP2015}. TOs are either Abelian or non-Abelian. For Abelian TOs, such as the toric code, the error-correction problem---which is to find the homology class of the physical error string with a given set of syndromes---has been well studied \cite{Dennis, Raussendorf, RMP2015}.\\
\indent In contrast, the error-correction problem of non-Abelian TOs is more difficult due to non-Abelian braiding statistics and nondeterministic fusion of anyon excitations \cite{Pachos_2012}. Previous work proved the existence of an error-correction threshold for non-Abelian TOs with perfect measurements and numerically demonstrated comparable thresholds for both Abelian and non-Abelian TOs using clustering decoders and renormalization group (RG) decoders \cite{BravyiHaah, WoottonLoss, Wootton_improved_HDRG, BurtonIsing, BurtonFibonacci, Verstraete, Wootton2016proof, HarringtonThesis, Dauphinais2017, HarringtonFibonacci} (for a passive error-correction approach, see Ref.~\onlinecite{chirame2024stabilizingnonabeliantopologicalorder}). However, unlike Abelian TOs, no optimal decoder is known for non-Abelian TOs. Moreover, although the existence of a threshold for continuous error correction in the presence of measurement errors has been proven for the special case of acyclic non-Abelian TOs, none of the previously studied decoders has inspired such a proof for the general non-Abelian case \cite{WoottonIsingproof, Dauphinais2017, davydova2025universalfaulttolerantquantum}.\\
\indent By definition, non-Abelian anyons, denoted by $a$, have multiple fusion channels
\begin{equation}
    a \; \times \; \Bar{a} \; = \; 1 \; + \; b \; + \; ... \; ,
    \label{fusion}
\end{equation}
leading to a quantum dimension $d_a > 1$. In contrast to the Abelian case, this indeterminacy of fusion implies that moving non-Abelian anyons requires a linear-depth circuit \cite{Beckman02, Shi19, Bravyi22, Liu22}, which cannot be achieved by local error channels. In this Letter, we utilize the information left behind by imperfect anyon strings \cite{Pablo1, Pablo2} to build `intrinsically heralded' decoders, as illustrated in Fig.~\ref{heralding}, which operate without the need for flag qubits \cite{Grassl, knill2004scalablequantumcomputationpresence, Kubica, Sahay, gu2023faulttolerantquantumarchitecturesbased, chirame2024stabilizingnonabeliantopologicalorder}. This uniquely non-Abelian property leads to improved thresholds compared to Abelian counterparts. Furthermore, we present an optimal decoder for non-Abelian TOs given perfect anyon syndromes, and discuss possible extensions to cases with measurement errors.\\
\indent While much of our discussion is general, we numerically confirm and illustrate our findings for the non-Abelian $D_4 \cong \mathbb Z_4 \rtimes \mathbb Z_2$ TO, for it has been recently realized in trapped ions \cite{D4preparation, Iqbal2024} and provides a resource for universal quantum computation \cite{davydova2025universalfaulttolerantquantum}. Our analytic and numerical analysis is worked out for the kagome lattice model of Ref.~\onlinecite{Yoshida}, with technical details in Ref.~\onlinecite{SM}. A key property of interest is that the $D_4$ TO has four charge anyons: three Abelian charge anyons transforming under the one-dimensional representations $s_{i = 1, 2, 3}$ of the $D_4$ gauge symmetry, and a non-Abelian anyon transforming under the two-dimensional representation $[2]$, with the fusion $[2] \times [2] = 1+ s_1 + s_2 + s_3$.\\
\indent \textit{Unheralded decoders of non-Abelian TO.} Quantum information encoded in the topological degeneracy of a TO is vulnerable to random unitary noise channels that drive the system into a mixed state. The goal of quantum error correction is to restore the system to its original state without corrupting the encoded information. Concretely, this requires the combined error and correction string to have trivial homology \cite{Dennis, RMP2015}.\\
\indent To extract information about the physical error, one measures the anyon content of the decohered mixed state, which serves as the error syndrome. Acyclic $a$ anyons, which do not appear among their own fusion outcomes on the right-hand side of Eq.~\eqref{fusion}, can be paired with antiparticles via shortest-length paths, a strategy known as minimum-weight perfect matching (MWPM) \cite{Dennis, WANG200331}. For instance, incoherent non-Abelian charge noise on the $D_4$ TO, corresponding to Pauli $\hat{X}$ noise on physical qubits, pair-creates $[2]$ anyons on the vertices of a honeycomb lattice \cite{Pablo2, chirame2024stabilizingnonabeliantopologicalorder}, whose MWPM threshold is $p_c = 0.15860(1)$, as shown by the black square in Fig.~\ref{phase_diagram}, coinciding with that of the toric code on the same lattice \cite{higgott2021pymatchingpythonpackagedecoding, MELCHERT20111828}. The threshold is discussed in the End Matter with numerical details provided in Ref.~\onlinecite{SM}.\\
\indent For cyclic $a$, an odd number of anyons may appear following pair-creation errors, rendering matching decoders inapplicable. As a result, previous work uses multiple iterations of perfect measurement and correction, even for one type of anyon \cite{WoottonLoss, Wootton_improved_HDRG, BurtonIsing, BurtonFibonacci, Verstraete}. However, these previous threshold estimates have relied exclusively on numerical results, and unlike the Abelian case, the error-correction problem has not been mapped to a statistical mechanics (stat-mech) model. Nevertheless, the existence of thresholds under arbitrary local error channels has been proven using an argument based on error clusters \cite{GACS198615, Gacs89, Gacs2001, HarringtonThesis, Wootton2016proof, Dauphinais2017}. As long as error correction prevents independently created clusters from growing and percolating into larger ones, logical errors are avoided when large clusters are sufficiently rare. Since anyons within a cluster will always fuse to the vacuum, regardless of their internal braiding and fusion, this argument applies to any TO. Indeed, previous studies have reported similar thresholds for clustering decoders and Harrington's RG decoder across Abelian TOs, the Ising anyon code, and the Fibonacci anyon code \cite{BravyiHaah, Wootton_toric_code, WoottonLoss, Wootton_improved_HDRG, BurtonIsing, BurtonFibonacci, Verstraete, HarringtonThesis, Dauphinais2017, HarringtonFibonacci}.\\
\indent \textit{Intrinsically heralded decoder for non-Abelian TOs.} While the proof and numerical determination of error-correction thresholds are valuable, it remains unclear how close these thresholds are to optimal. Moreover, although previously studied decoders have achieved comparable thresholds for Abelian and non-Abelian TOs, they have neither utilized the properties of the TOs to inform decoding nor provided insight into whether the non-Abelian nature offers any advantage for error correction.\\
\indent A distinctive feature of non-Abelian anyons is that their movement requires a linear-depth quantum circuit. In contrast, physical noise, such as the single-qubit Pauli channel, corresponds to finite-depth operations and cannot generate exact non-Abelian error strings beyond pair creation on neighboring sites. As a result, along a string of physical error qubits, non-Abelian anyons are created at the endpoints, while superpositions of vacuum and intermediate anyons are present along the path, reflecting the unresolved non-Abelian fusion channels \cite{Pablo1, Pablo2}. The syndromes of both non-Abelian and intermediate anyons can be extracted by measuring the commuting projectors of the Hamiltonian that defines a fixed-point TO
\begin{equation}
    H\;=\;\sum_iA_i,
    \label{commuting_projector}
\end{equation}
where the eigenvalues of $A_i$ are $0$ or $1$ \cite{KITAEV20032, LevinWen}. When the only errors are incoherent pair creations of non-Abelian $a$ anyons, the detection of other types of intermediate anyons (e.g., $b$ in Eq.~\eqref{fusion}) provides a clear signature that the error string has acted on the site. Therefore, we can modify unheralded decoders to require the error-correction string to pass through all intermediate anyons. This constitutes the intrinsic heralding that informs and enhances error correction, as shown in Fig.~\ref{heralding}.\\
\indent For the $D_4$ TO under incoherent non-Abelian charge noise, the intrinsically heralded MWPM decoder selects the shortest correction string that traverses all measured Abelian charges, which also reside on the vertices of the same honeycomb lattice, as shown in Fig.~\ref{phase_diagram}. By modifying the edge weights of MWPM to $w=1-n_e K$, where $n_e$ is the number of $e$ charges connected to the edge and heralding strength $K$ is a sufficiently large constant effectively acting as infinity, this achieves a threshold of $p_c = 0.20842(2)$ (yellow symbol in Fig.~\ref{phase_diagram}) for logical errors arising from homologically nontrivial $[2]$ anyon loops, a significant increase compared to that of the unheralded ($K = 0$) MWPM decoder, which has $p_c = 0.15860(1)$ (black square), without needing flag qubits. For details, see Ref.~\onlinecite{SM}.\\
\indent While the intrinsically heralded decoder improves the identification of the physical error string, full recovery of the encoded quantum state still requires multiple rounds of measurement and error correction, since the nondeterministic fusion of non-Abelian anyons \cite{WoottonLoss, Wootton_improved_HDRG, BurtonIsing, BurtonFibonacci, Verstraete} can create new anyons along the correction string. Because minimum-weight corrections do not increase the size of independently created error clusters \cite{Wootton2016proof}, they remain nonpercolating as long as the physical error parameters lie within the error-correcting phase. In other words, the threshold against incoherent pair-creation errors of a single anyon type is set by the first application of intrinsically heralded minimum-weight decoding, as discussed in detail in the End Matter. Confirming this in practice, when accounting for logical errors arising from both the correction of non-Abelian $[2]$ charges and the subsequent MWPM of Abelian charges, the $D_4$ thresholds under $[2]$ anyon noise are found to be $p_c = 0.1585(3)$ for unheralded $[2]$ MWPM and $p_c = 0.2084(5)$ for intrinsically heralded MWPM, both within one standard deviation of the thresholds considering only nontrivial $[2]$ loops.\\
\indent More generally, a single round of finite-depth correction suffices to annihilate all acyclic $a$ anyons generated by incoherent pair-creation noise. In contrast, annihilating cyclic $a$ anyons along the minimal error tree requires resources equivalent to perfect measurement and correction using linear-depth circuits repeated for a logarithmic number of iterations in the linear system size, since each iteration fuses anyons within each connected component of the tree, reducing their number by at least half.\\
\indent \textit{Optimal Decoding for non-Abelian TOs.} Instead of correcting along the shortest error string, a decoder may consider all admissible error configurations. To the best of our knowledge, this has thus far been explored only in the context of optimal decoding of Abelian TOs. The seminal work by Dennis \textit{et al.} \cite{Dennis} on the toric code can be interpreted as calculating the probability of homology class $h$ given anyon syndromes $\bm a$:
\begin{equation}
    P(h|{\bm{a}}) = \sum_{E\in h} P(E|{\bm{a}}) \propto \sum_{E\in h,\partial E={\bm{a}}} \left(\frac{p}{1-p}\right)^{|E|},
    \label{TC}
\end{equation}
where two error paths $E$ are said to be in the same homology class if they can be continuously deformed into one another. The error-correcting phase is characterized by the dominance of a single class in the infinite code distance limit, which determines the optimal decoding strategy. This approach can also be applied to acyclic non-Abelian $a$ anyons. For the $D_4$ TO under $[2]$ anyon noise, using only the $[2]$ syndromes yields a threshold no better than $p_c = 0.1642(3)$, corresponding to the multicritical point of the random-bond Ising model on the triangular lattice \cite{Nishimori, Honecker, Merz, Lessa, Queiroz09, RBIMtriangular}, shown in Fig.~\ref{phase_diagram} as the black star, with details provided in the End Matter.\\
\indent To find an optimal decoder for non-Abelian TOs under arbitrary noise, we consider decoding conditioned on the full syndrome $\bm{s} = \{\bm{a}, \bm{b}, \dots\}$. Reliable inference of the physical error $E$ is only possible if the conditional probability $P(E | \bm{s})$ (for fixed $\bm{s}$) lies in a `short string' phase. Although $P(E|\bm{s})$ is a challenging quantity to calculate, we can make it tractable by applying Bayes' theorem:
\begin{equation}
    P(E|\bm{s}) = \frac{P(\bm{s}|E)P(E)}{P(\bm{s})} \propto P(\bm{s}|E)P(E)
    \label{Bayes}
\end{equation}
While $P(E)\propto (\frac{p}{1-p})^{|E|}$ represents the probability of physical single-qubit errors occurring along the string $E$, $P(\bm{s}|E)$ accounts for the probabilistic collapse of superpositions over non-Abelian fusion channels into a specific set of intermediate anyons along $E$. This quantity is given by the expectation value of the anyon projectors evaluated on the corrupted quantum state $\ket{E}$:
\begin{equation}
    P(\bm{s}|E)=\bra{E} \prod_{i} \left[ (1-\lambda_i)(1-A_i) + \lambda_iA_i \right] \ket{E},
    \label{Probability}
\end{equation}
where $\lambda_i\in\{0,1\}$ is the measurement outcome of projector $A_i$. The optimal threshold is then identified with a `string proliferation' phase transition of the quenched-disorder stat-mech model, where syndrome $\bm s$ is the `disorder' variable, and $P(E|\bm s)$ is a stat-mech model on the configuration space of errors. Since expectation values are insensitive to scaling stat-mech weights by an arbitrary function $f(\bm s)$, we can choose analytic weights $ P(\bm{s}\cap E) = P(\bm{s}|E)P(E)$, such that the partition function $\mathcal Z_{\bm{s}} = \sum_E P(\bm{s}\cap E)$ equals the quenched disorder probability, $Z_{\bm{s}} = P(\bm{s})$, known as the Nishimori property \cite{Nishimori81, Gruzberg01, Dennis, Zhu23, Lee22, Fan24, Lee23, Putz25, Nahum25}.\\
\indent If the noise model corresponds to incoherent pair-creation in the anyon basis, the stat-mech model gives an explicit optimal decoder that corrects with the error string in the most likely homology class, which maximizes $P(h|\bm{s}) \propto \sum_{E\in h} P(\bm{s}|E)P(E)$. More generally, if the noise model creates coherence in the anyon basis, the optimal threshold given by the stat-mech model can be achieved by decomposing the physical error into effective incoherent anyon strings and decoding with anyon strings in the most likely homology class for all anyons, as detailed in the End Matter. Although nondeterministic fusion can create new anyons after each correction, the initial step determines the true bottleneck and thus sets the threshold $p_c$, as previously reasoned for MWPM decoders and argued in the End Matter.\\
\indent The above construction of optimal decoders applies generally to any non-Abelian TO under arbitrary noise. However, a key challenge in determining the optimal threshold lies in efficiently evaluating $P(h|\bm{s})$, which in turn depends on the efficient computation of $P(\bm{s}|E)$. In Ref.~\onlinecite{SM}, we design a Monte Carlo protocol to sample contributing terms in $P(h|\bm{s})$ for the $D_4$ TO under $[2]$ noise \cite{ParisenToldin_2006, Hasenbusch_2007, Hasenbusch_2007_2, ParisenToldin2009}, with example calculations of $P(\bm{s}|E)$ shown in Fig.~\ref{phase_diagram}, and find the optimal threshold to be $p_c = 0.218(1)$, corresponding to the orange star on the Nishimori line $\beta = \ln\sqrt{\frac{1-p}{p}}$ (dashed) in Fig.~\ref{phase_diagram}. This shows that the threshold of intrinsically heralded MWPM is close to optimal. Furthermore, in Ref.~\onlinecite{SM}, we give a stat-mech model for the $D_4$ TO under arbitrary Pauli noise, leaving the question of its efficient sampling to future work.\\
\begin{figure}[t]
\includegraphics[width=0.9\columnwidth]{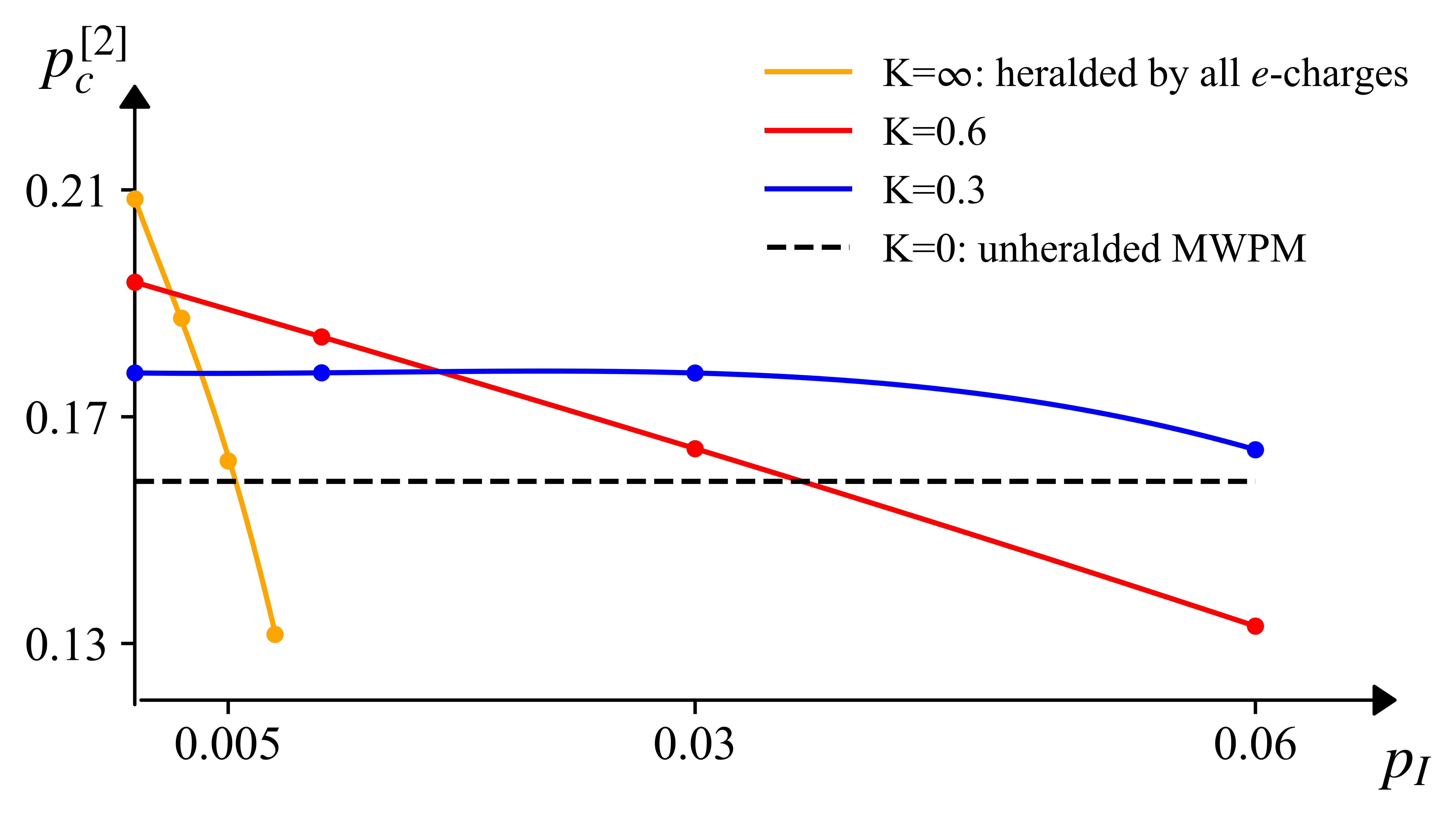}
\caption{\textbf{Stability of heralding.} Error-correction threshold of non-Abelian charges in the $D_4$ TO as a function of the intermediate Abelian charge pair-creation rate $p_I$ for different heralding strength $K$. Solid lines show the $[2]$ anyon thresholds $p_c^{[2]}$ for intrinsically heralded ($K \neq 0$) MWPM decoders, while the black dashed line indicates the unheralded ($K = 0$) MWPM threshold. Decreasing $K$ reflects lowered confidence in the correctness of heralding and improves threshold at higher $p_I$ at the expense of lowering threshold at small $p_I$. For $K = 0.3$, the advantage from intrinsic heralding persists up to $p_I \approx 6\%$, the threshold for Abelian charge noise alone.}
\label{Z_tolerance}
\end{figure}
\indent \textit{Stability of heralding.} When intermediate anyons, such as the Abelian charges in the $D_4$ TO, are pair-created by the error channel, they can lead to incorrect heralding of non-Abelian anyons. Focusing on the $[2]$ anyon threshold $p_c^{[2]}$, the orange solid line in Fig.~\ref{Z_tolerance} shows that the threshold improvement for non-Abelian charges over the unheralded MWPM decoder (black dashed line) persists only up to an intermediate anyon (Abelian charge) pair-creation rate of $p_I \approx 0.5\%$ when all measured intermediate anyons are used for heralding ($K = \infty$). However, reducing the heralding strength $K$, which reflects lowered confidence in the correctness of the heralding, improves the non-Abelian threshold at higher $p_I$ at the expense of a lower threshold at small $p_I$, as illustrated by the red solid line ($K = 0.6$) relative to the yellow one ($K = \infty$). Furthermore, with $K = 0.3$ (blue solid line), the improvement persists up to $p_I = 6\% \approx p_c^I$, where $p_c^I$ is the threshold for Abelian charge noise in the absence of non-Abelian noise. This demonstrates that intrinsically heralded decoders can be adapted to remain effective across a wide range of error regimes.\\
\indent \textit{Measurement errors.} In practice, measurement errors necessitate repeated syndrome measurements to identify faulty outcomes and continuously correct physical errors \cite{Dennis, HarringtonThesis, Wootton_improved_HDRG, Dauphinais2017}. Even when the number of repeated measurements is chosen adaptively based on confidence in the measurement outcomes, or `just in time' \cite{bombinJIT, BrownJIT, ScrubyJIT, davydova2025universalfaulttolerantquantum}, reliably identifying measurement errors may still require many repetitions. However, instead of commuting-projector measurements, non-Abelian anyon syndromes can be extracted by measuring the corresponding terms in a quasistabilizer Hamiltonian whose non-Pauli components commute only within the ground-state subspace \cite{Yoshida, chirame2024stabilizingnonabeliantopologicalorder}. In this case, sudden fluctuations of intermediate anyons provide reliable signatures of measurement errors, effectively acting as a timelike heralding. This uniquely non-Abelian approach could be exploited to improve the efficiency of measurement error detection.\\
\indent Furthermore, although valuable for proofs of fault tolerance and broadly applicable to any TO, existing RG and `just-in-time' decoders do not themselves provide efficient correction schemes once errors are identified; instead, they must be integrated with other decoders, such as clustering or MWPM, as discussed in Ref.~\onlinecite{davydova2025universalfaulttolerantquantum}. In this context, intrinsic heralding serves as a valuable complement: just as it enhances MWPM decoding under perfect measurements, intrinsic heralding in both space and time exploits the properties of the TO to improve error identification and enable more efficient removal of error clusters once identified.\\
\indent For acyclic anyon $a$, identifying the pair-creation and measurement error strings reduces to a matching problem, which no longer requires `just-in-time' techniques. This problem can be mapped onto a three-dimensional stat-mech model, similar to that in Ref.~\onlinecite{Dennis}, where the additional dimension represents time. Although correcting a single anyon type $a$ protects a classical bit rather than a qubit, the problem is nevertheless well-defined. Moreover, successful correction of $a$ is a necessary condition for correcting all anyons and thus for protecting the encoded qubit, and its threshold is likely to be the bottleneck, especially when errors are biased toward $a$. In Ref.~\onlinecite{SM}, we present an example of identifying measurement errors and suggest a construction of an intrinsically heralded matching decoder in both space and time for the $[2]$ anyon of the $D_4$ TO.\\
\indent \textit{Outlook.} A natural next step is to apply intrinsic heralding to cyclic non-Abelian anyons that self-herald their error correction. Examples of this kind include Fibonacci anyons and $S_3$ TO, which support universal quantum computation \cite{Mochon04, Lo25, Chen2025, Minev2025}. As for optimal decoding under arbitrary local noise, efficient sampling of the identified stat-mech model remains an open problem. More generally, it would be interesting to characterize the phase diagram of these novel stat-mech models, where the quantum dimension of the proliferated anyon enters directly into the Boltzmann weights and thus affects the resulting phase diagram. Additional insight may be drawn from Ref.~\onlinecite{Nahum25}, which follows a similar Bayesian approach. Other interesting directions include investigating our optimal strategy under coherent noise \cite{Bravyi2018, Venn23, Lavasani24, Bao24, Behrends24, Behrends24b, Hauser25} and, since our optimality is conditioned on measuring anyon syndromes, exploring whether alternative measurement bases can further improve non-Abelian thresholds. Recent works \cite{Pablo1, Pablo2} suggest a potential threshold as high as $p_c \approx 0.5$ for pure non-Abelian noise in certain cases, although no explicit decoder has been constructed.\\
\indent In the presence of measurement errors, a quantitative determination of the improvement afforded by space-time heralding has yet to be carried out. Although this improvement is expected to persist under unstable space-time heralding, its magnitude and robustness outside regimes biased toward non-Abelian errors are open questions. Furthermore, error information can be hidden in the internal degrees of freedom of non-Abelian anyons, making it inaccessible to anyon syndrome measurements. This suggests going beyond occupation measurements, for example by using adaptive measurement bases, and incorporating intrinsic heralding into local cellular-automata decoders \cite{localautomaton}.\\
\indent \textit{Note added.} We recently noted a preprint that mentioned heralded decoding of non-Abelian $S_3$ TO, although a specific decoder was not proposed \cite{sajith2025noncliffordgatesstabilizercodes}.\\
\indent \textit{Acknowledgments.}
The authors thank Yimu Bao for an enlightening discussion that helped us generalize our results to the case of coherent errors.
D. J. would like to acknowledge Sanket Chirame for his explanation of the stabilizer tableaux for simulating the $D_4$ TO, Dr.~Trung Nguyen for assistance with the Midway computing cluster, Shu Shi for discussions on classical algorithms and data structures, Professor Werner Krauth and Dr.~Nils Strand for discussions on Monte Carlo methods, and Dr.~Ramanjit Sohal for discussions on topological orders. D. J. thanks participants of the 2025 APS March Meeting for discussion and feedback on the reporting of our results \cite{MMtalk}.
P. S. would like to acknowledge the participants of the Perimeter Institute's `Non-Abelian Anyon' summer camp for their valuable feedback, Olexei Motrunich for pointing out relevant literature on the reweighting technique, and the support from the Caltech Institute for Quantum Information and Matter, an NSF Physics Frontiers Center (NSF Grant No. PHY-1733907), and the Walter Burke Institute for Theoretical Physics at Caltech. R. V. thanks Henrik Dreyer, Sagar Vijay, and Ashvin Vishwanath for discussion and encouragement.
This work is supported in part by the DARPA MeasQuIT program and completed using resources provided by the University of Chicago's Research Computing Center.  

\bibliographystyle{apsrev4-2}
\bibliography{references}

\onecolumngrid
\begin{center}
\vspace{1em}
\textbf{End Matter}
\end{center}
\twocolumngrid

\indent After physical errors act on the quantum information initially encoded in $\rho_0=\ket{\psi_0}\bra{\psi_0}$, the state is a decohered mixed state $\sum_E P(E)\hat{E}\rho_0 \hat{E}^\dagger$, where $\hat{E}$ denotes a unitary error occurring with probability $P(E)$. Anyon measurements collapse the density matrix, making it block-diagonal in the anyon basis,
\begin{equation}
    \rho = \bigoplus_{\bm s} P(\bm s) \; \rho_{\bm s},
    \label{density_matrix0}
\end{equation}
where $\rho_{\bm s}$ is the normalized density matrix associated with the block labeled by syndromes $\bm s$. Each block generally has dimensions larger than one, arising from the internal degrees of freedom of non-Abelian anyons or from distinct homology classes on topologically nontrivial manifolds. As the internal degrees of freedom are not accessible through measurements, we henceforth assume they have been traced out.\\
\indent Starting from Eq.~\eqref{density_matrix0}, we rederive Eq.~\eqref{Bayes} and verify the validity of the Bayesian inference. On an infinitely large lattice with only trivial homology, $\rho$ is diagonal in the anyon basis after tracing over internal degrees of freedom. As discussed in the main text, the error-correction threshold can be identified with the phase transition of a quenched-disorder stat-mech model whose partition function, for each syndrome measurement outcome $\bm s$ and satisfying the Nishimori property, is given by $Z_{\bm s}=P(\bm s)=\mathrm{Tr}(\Pi_{\bm s}\rho\Pi_{\bm s})$, where $\Pi_{\bm s}$ is the projector onto the block labeled by $\bm s$. Expanding over the configuration space of physical errors yields 
\begin{align}
    Z_{\bm s}&=\sum_E P(E)\mathrm{Tr}(\Pi_{\bm s}\hat{E}\rho_0 \hat{E}^\dagger\Pi_{\bm s}) \nonumber\\
    &=\sum_E P(E) \bra{\psi_0}\hat{E}^\dagger\Pi_{\bm s}\hat{E}\ket{\psi_0}
    \label{infinite_lattice}
\end{align}
thus recovering Eq.~\eqref{Bayes} using Eq.~\eqref{Probability} in the main text. Averaging over the `disorder' variable $\bm s$, the free energy is $\langle\beta F\rangle_{\bm s}=-\sum_{\bm s}P(\bm s)\ln{Z_{\bm s}}=-\sum_{\bm s}P(\bm s)\ln{P(\bm s)}$. A singularity in the averaged free energy signals a phase transition of the stat-mech model \cite{Dennis}, which, in some cases, can be conveniently detected by sampling order parameters associated with the physical error configurations $E$.\\
\indent Quantum information can be encoded using TOs defined on manifolds with nontrivial homology. Within each block $\rho_{\bm s}$, the states are labeled by the homology classes of all anyon types, $h=\{h_a,\;h_b,\;\ldots\}$. Here, anyon strings terminating at the same set of syndromes are said to belong to the same homology class if they can be smoothly deformed into one another, or equivalently if the sum of any pair of strings forms only homologically trivial loops. For physical errors $E$ that are \emph{incoherent} in the anyon basis, their homology classes $h_E$ are well-defined, with probabilities given by the partition function
\begin{equation}
    Z_{\bm s,h_E}=P(\bm s,h_E)=\sum_{E\in h_E} P(E)\mathrm{Tr}(\Pi_{\bm s}\hat{E}\rho_0 \hat{E}^\dagger\Pi_{\bm s}).
    \label{variable_E}
\end{equation}
In the error-correcting phase, the probability of a single homology class approaches $1$ in the limit of infinite code distance, while those of all other classes vanish. Error correction can then be performed by applying any $\hat{E}$ operator in the dominant class \cite{Dennis}.\\
\indent In general, physical errors are coherent in the anyon basis and can be decomposed as $\hat{E}=\sum_\epsilon \omega(\epsilon|E)\hat{\epsilon}$, where $\hat{\epsilon}$ consists of \emph{incoherent} anyon strings that are consistent with syndromes $\bm s$ and have support contained within that of $\hat{E}$. This leads to ill-defined homology $h_E$, which therefore cannot be directly used for error correction. Instead, $\rho_{\bm s}$ has matrix elements
\begin{equation}
    \bra{h_L}\rho_{\bm s}\ket{h_R}=\frac{1}{P(\bm s)}\sum_E \sum_{\substack{\epsilon_L\in h_L \\ \epsilon_R\in h_R}}P(E)\omega(\epsilon_L|E)\omega^*(\epsilon_R|E),
    \label{reduced_rho_elements}
\end{equation}
where $h_{L,R}$ are the homology classes of all anyon types. While the diagonal entries of $\rho_{\bm s}$ are real and positive, the off-diagonal elements are generally complex and nonzero.\\
\indent To perform optimal decoding, one applies anyon strings in the most likely homology class, corresponding to $\arg\max_h\bra{h}\rho_{\bm s}\ket{h}$ \cite{Bao24}. The error-correcting phase of the decoder is defined as the region in the parameter space of the physical errors where, in the limit of infinite code distance, a unique homology class $h_i$ dominates with $\bra{h_i}\rho_{\bm s}\ket{h_i}=1$, while all other homology classes satisfy $\bra{h_j\neq h_i}\rho_{\bm s}\ket{h_j\neq h_i}=0$. This definition yields the same phase transition, or equivalently optimal threshold, as the singularity in $\langle\beta F\rangle_s$ on the infinitely large lattice \cite{Dennis, WANG200331}. When a diagonal element of a density matrix is zero, the requirement of positive semi-definiteness forces the entire corresponding row and column to vanish. Consequently, in the error-correcting phase in the infinite code distance limit, $\rho_{\bm s}$ has a single diagonal element equal to one, with all other entries vanishing, which implies that error correction succeeds with certainty and the state is restored to $\rho_0=\ket{\psi_0}\bra{\psi_0}$.\\
\indent For the error correction of the $D_4$ TO defined on a torus under non-Abelian charge noise, the syndromes $\bm s$ consist of non-Abelian $[2]$ charges and intermediate anyons $I$. Consequently, the states in $\rho_{\bm s}$ are labeled by the homology classes of $[2]$ and all intermediate anyons, $\{h_{[2]},\; h_I\}$. The physical errors are incoherent among states labeled by $h_{[2]}$ and coherent among states labeled by $h_I$. As a result, $\rho_{\bm s}$ is block-diagonal in $h_{[2]}$,
\begin{equation}
    \rho_{\bm s} = \bigoplus_{h_{[2]}} P(h_{[2]})\rho_{\bm s,h_{[2]}},
    \label{variable_incoherent}
\end{equation}
where each normalized block $\rho_{\bm s,h_{[2]}}$ consists of states labeled by $h_I$ and has nonzero off-diagonal elements.\\
\indent Furthermore, the error model requires the intermediate anyon strings to lie along the $[2]$ strings in $\hat{\epsilon}_{L,R}$. This implies that the homology class $h_E$ and $h_{[2]}$ are equivalent, and the probabilities $P(\bm s,h_E)$ in Eq.~\eqref{variable_E} are the same as $P(h_{[2]})$ in Eq.~\eqref{variable_incoherent}. As a result, below the same threshold, a unique homology class $h_E$ and the corresponding block $\rho_{\bm s,h_{[2]}}$ dominates with probability one in the limit of infinite code distance. 
Because intermediate anyon strings are required to lie along the $[2]$ strings, this definite homology class of the $[2]$ component in $\hat{\epsilon}_{L,R}$ ensures that the homology class of intermediate anyons, $h_I$, is also unique.
Consequently, $\rho_{\bm s,h_{[2]}}$, and hence $\rho_{\bm s}$ in Eq.~\eqref{variable_incoherent}, take the desired form, containing a single nonzero diagonal element equal to $1$. The fact that a unique $h_{[2]}$ ensures a unique $h_I$ indicates that the threshold for intermediate anyons created along $[2]$ strings lies above that of $[2]$ anyons themselves, a conclusion corroborated by our numerical results. Thus, the decoders defined in Eqs.~\eqref{variable_E} and~\eqref{variable_incoherent} exhibit the same optimal threshold, set by that of the $[2]$ anyons.\\
\indent Optimal decoding can then be performed using the diagonal elements of $\rho_{\bm s}$, computed via Eq.~\eqref{variable_incoherent}, with anyon string operators employed for error correction. However, to avoid the use of linear-depth non-Abelian anyon strings, we first apply the same finite-depth operators appearing in the physical errors of the most likely homology class, identified using either Eq.~\eqref{variable_E} or~\eqref{variable_incoherent}, to pair-annihilate the $[2]$ charges, while deferring the pairing of intermediate anyons until after the $[2]$ anyons have been corrected. Since the above analysis establishes that the correction of $[2]$ anyons has the same threshold as the optimal decoders in Eqs.~\eqref{variable_E} and~\eqref{variable_incoherent}, it suffices to account for the additional coherent errors among intermediate anyons that arise along the finite-depth error-correction strings used for the pair-annihilation of $[2]$ anyons to prove that our two-step decoding protocol is optimal.\\
\indent After applying these strings, the density matrix in Eq.~\eqref{variable_incoherent} becomes $\rho_{\bm s}^\prime = \bigoplus_{h_{\circ}} P(h_{\circ})\rho_{\bm s,h_{\circ}}^\prime$, where $h_{\circ}$ denotes the homology class of the closed loops formed by combining the physical error strings with the error-correction strings. $P(h_{\circ})$ corresponds one-to-one to $P(h_{[2]})$, since the error-correction strings are also incoherent for $[2]$ anyons. The normalized block $\rho_{\bm s,h_{\circ}}^\prime$ contains both the coherence in $\rho_{\bm s,h_{[2]}}$ and the additional coherence introduced by the error-correction strings. Below the threshold of the decoders defined in Eqs.~\eqref{variable_E} and~\eqref{variable_incoherent}, and in the limit of infinite code distance, the trivial homology class in $h_{\circ}=1$ dominates with $P(h_{\circ}=1)=1$. By the same reasoning as before, loops in the trivial class contain no homologically nontrivial components. Consequently, the homology class of intermediate anyons is uniquely fixed and $\rho_{\bm s,h_{\circ}}^\prime$ contains a single diagonal element equal to one, with all other entries vanishing. Hence, the error-correcting phase of the decoders defined in Eqs.~\eqref{variable_E} and~\eqref{variable_incoherent} guarantees the correction of coherence among intermediate anyons. This concludes our proof that the two-step decoding protocol against non-Abelian $[2]$ charge noise in the $D_4$ TO is indeed optimal.\\
\indent Therefore, to numerically determine the optimal threshold of the $D_4$ TO (orange star in Fig.~\ref{phase_diagram}), we sample the order parameter associated with the physical error $E$ in Eq.~\eqref{infinite_lattice}. The physical error string $E$ and the disorder configuration $\bm{s}$ are first randomly generated using a disjoint-set union data structure, ensuring that the resulting parity and entanglement structure of $I$ are consistent with those derived from the stabilizer formalism \cite{chirame2024stabilizingnonabeliantopologicalorder, SM}. Then, the summation in Eq.~\eqref{infinite_lattice} can be rewritten as the partition function of a stat-mech model of classical Ising spins $\{\sigma\}$,
\begin{gather}
    \mathcal{Z}_{\bm{s}} \propto \sum_{\text{allowed}\;\{\sigma\}}2^Ce^{-\beta H}, \;\;
    H = - \sum_{\langle i,j \rangle} \eta_{ij}(E)\tilde{w}_{ij}(\bm{s})\sigma_i\sigma_j,\nonumber\\
    \tilde{w}_{ij}(\bm{s})=1-n_I K+ \frac{2-n_{\bm {[2]}}}{2} \frac{\ln{2}}{2\beta},
\end{gather}
where $\eta_{ij} = -1$ along the error string $E$ and $\eta_{ij} = 1$ elsewhere, $n_{I/\bm{[2]}}\in\{0, 1, 2\}$ is the number of intermediate or $\bm{[2]}$ anyons connected to the edge, and $C$ is the number of parity constraints on intermediate anyons. The fluctuations of the Ising spins ${\sigma}$ generate all physical error strings satisfying $\partial E = \bm{[2]}$. These error strings are required to pass through all intermediate anyons, a condition enforced by introducing an arbitrarily large constant $K$, and to satisfy the set of constraints $C$, which are verified using a graph search algorithm and assigned zero weight if violated. The moments of the average magnetization of ${\sigma}$ are sampled using local-update Monte Carlo simulations weighted by $e^{-\beta H}$ and reweighted by $2^{C}$ to account for nonlocal constraints that are computationally costly to evaluate at each step. Binder cumulants are then computed from these moments, after averaging over the disorder variable $\bm{s}$, to determine the optimal threshold through finite-size scaling analysis. Further details can be found in Ref.~\onlinecite{SM}.\\
\indent Lastly, we discuss how to construct the unheralded optimal decoder that pair-annihilates incoherently created non-Abelian $[2]$ charges without the knowledge of intermediate anyons. The decohered mixed state after introducing physical errors, on an infinitely large lattice with only trivial homology, takes the block-diagonal form
\begin{equation}
    \rho=\bigoplus_{\bm {[2]}}\rho_{\bm {[2]}},
\end{equation}
where each block $\rho_{\bm {[2]}}$ is spanned by states labeled by anyon configurations, $\ket{\bm s =\{\bm {[2], I}\}}$, and remains unchanged upon the measurement of only $[2]$ anyons. Furthermore, the diagonal elements of $\rho$ correspond to
\begin{align}
    \bra{\bm {[2], I}}\rho\ket{\bm {[2], I}}&=\sum_{E} P(E)\mathrm{Tr}(\Pi_{\bm {[2], I}}\hat{E}\rho_0 \hat{E}^\dagger\Pi_{\bm {[2], I}}) \nonumber\\
    &=\sum_{E} P(E)P(\bm {[2], I}|E).
\end{align}
Since the unheralded decoder does not measure or make use of the syndrome information associated with intermediate anyons, the subspace within each $\rho_{\bm {[2]}}$ block corresponding to different intermediate anyon configurations $\bm I$ but fixed non-Abelian syndromes $\bm {[2]}$ can be traced out, yielding a reduced diagonal density matrix
\begin{equation}
    \tilde{\rho}=\sum_{\bm {[2]}} \ket{\bm {[2]}}\bra{\bm {[2]}}\sum_{E} P(E) \sum_{\bm I} P(\bm {[2], I}|E),
\end{equation}
the diagonal elements of which correspond to
\begin{align}
    P(\bm{[2]})&=\sum_{E} P(E) \sum_{\bm I} P(\bm {[2], I}|E) \nonumber\\
    &=\sum_{E} P(E) P(\bm {[2]}|E)=\sum_{\partial E=\bm {[2]}} P(E).
\end{align}
Setting $Z_{\bm{[2]}}=P(\bm{[2]})$ recovers the stat-mech model for the optimal toric code decoder \cite{Dennis}, identical to that for the optimal unheralded decoding of non-Abelian $[2]$ charges. The unheralded MWPM decoder is recovered by replacing the sum $\sum_{\partial E=\bm {[2]}}$ with its largest term.

\appendix
\onecolumngrid

\newcounter{suppappendix}
\renewcommand{\thesuppappendix}{\Alph{suppappendix}}

\newpage
\refstepcounter{suppappendix}
\setcounter{equation}{0}
\renewcommand{\theequation}{\thesuppappendix\arabic{equation}}
\renewcommand{\theHequation}{supp.\thesuppappendix.\arabic{equation}}
\section{Stabilizer Formalism for Error Correction in the $D_4$ Topological Order}
\label{stabilizer}
\indent We demonstrate the improved error-correction threshold of our intrinsically heralded decoder using the $D_4$ TO realized in the kagome lattice model of Ref.~\onlinecite{Yoshida}. In this Appendix, we consider a single color of Pauli $\hat{X}$ errors, which create pairs of $m$-anyons at the endpoints of error strings and leave a superposition of vacuum and $e$-anyons along their paths. Without loss of generality, we take the Pauli $\hat{X}$ errors to occur on red qubits, resulting in intermediate $e$ charges on blue and green stars. We analyze the error correction for the $D_4$ TO under this error channel and derive constraints on the measurement outcomes of intermediate Abelian anyons relevant to our numerical simulations. Additionally, we discuss the calculation of $P(\bm{s}|E)$ in Eq.~\eqref{Probability} of the main text.
\subsection{$D_4$ TO on three-colorable kagome lattice}
\indent The $D_4$ TO can be defined on a three-colorable kagome lattice by the quasistabilizer Hamiltonian
\begin{equation}
    H_{D_4}
    \;=\;
    -\sum_{s\in \{\textrm{\ding{65}}\}}A_s
    \;-\;
    \sum_{t\in \{\triangleright,\,\triangleleft\}} B_t,    
    \label{D4}
\end{equation}
where the star and triangle operators are defined as
\begin{equation}
    \vcenter{\hbox{\includegraphics[height=5.5em]{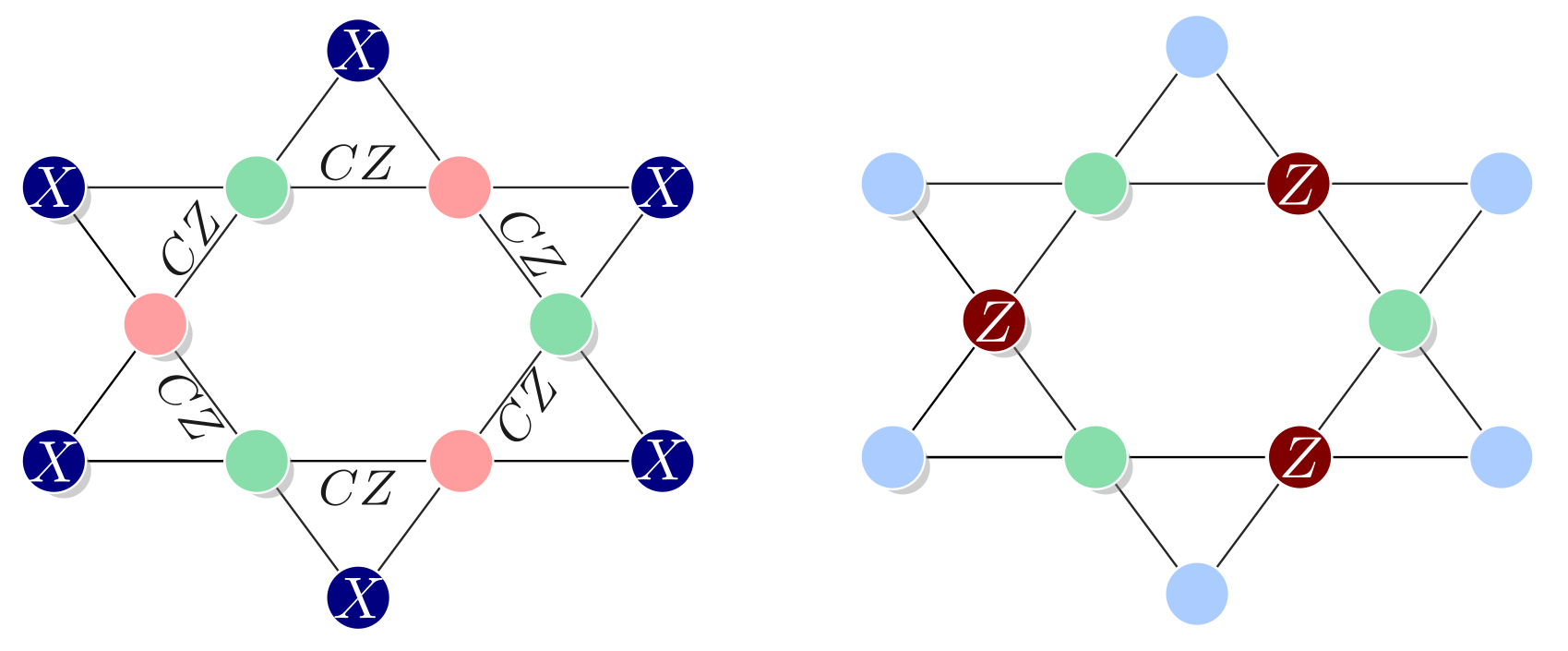}}}
    \label{Hamiltonian_terms}
\end{equation}
Each star and triangle term in the Hamiltonian has eigenvalues $\pm1$ \cite{Yoshida, Iqbal2024, D4preparation}. In this representation, the TO is a $\mathbb Z_2^3$ gauge theory `twisted' to become non-Abelian \cite{Yoshida}. Using that terminology, the violations of the star operators create Abelian $e$ charges, whereas excited triangle operators correspond to non-Abelian $m$-fluxes. The anyons also carry a color label inherited from the lattice. The fusion rules are $m_R\times m_R = 1 + e_B + e_G + e_B e_G$, $e_B\times e_B = 1$, $e_G\times e_G = 1$, and their color permutations. Notably, the non-Abelian anyons fuse to Abelian ones, which is an example of being acyclic \cite{Dauphinais2017}. Besides multiple fusion channels for the $m$-anyons, the non-Abelian nature of the TO is also manifested in the following commutation relation
\begin{equation} \label{commutation}
    \vcenter{\hbox{\includegraphics[height=5.5em]{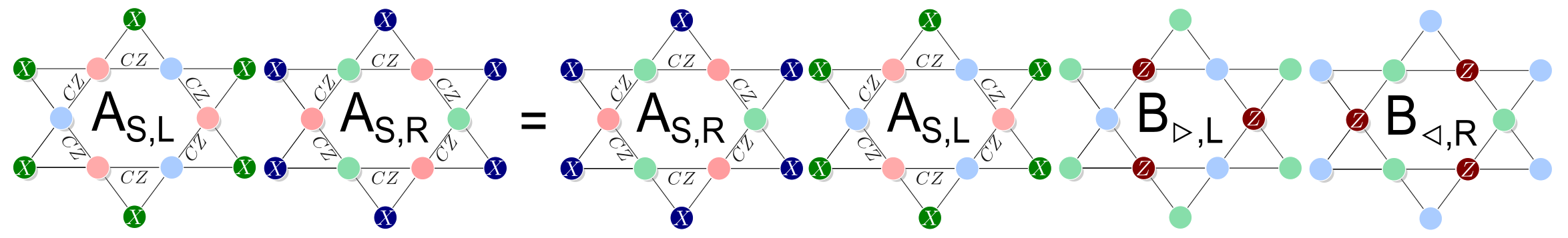}}}
\end{equation}
where $L$ and $R$ are adjacent stars. While all Hamiltonian terms commute in the 22-fold degenerate ground space, the presence of an $m$-anyon causes the star operator whose support overlaps with the anyon to no longer commute with three of its neighboring stars. The measurement of such star operators should therefore be avoided.\\
\indent Remarkably, this `twisted $\mathbb Z_2^3$ gauge theory' realizes the same topological order as the $D_4$ quantum double \cite{Yoshida}. If we use the language of $D_4$ gauge theory, then the non-Abelian charge anyon corresponds to $m_c$ for an arbitrary choice of color. We choose the convention $[2] = m_R$ going forward. Moreover, the three Abelian charges $s_{i = 1, 2, 3}$ of the $D_4$ gauge theory then correspond to $e_B$, $e_G$ and $e_B e_G$. Hence, the above fusion rule agrees with the one claimed in the main text. For a full dictionary between the anyons of $D_4$ gauge theory and the twisted $\mathbb Z_2^3$ gauge theory, we refer to the appendix of Ref.~\onlinecite{Iqbal2024}. Due to the aforementioned equivalence, we will freely switch between talking about the non-Abelian charge anyon $[2]$ and the non-Abelian flux anyon $m_R$.\\
\indent Alternatively, the $D_4$ TO realized in the kagome lattice model can be defined by a commuting-projector Hamiltonian
\begin{equation}
\label{Proj_dressing}
    H_{D_4}
    \;=\;
    \sum_{s\in \{\textrm{\ding{65}}\}}A_s^p
    \;+\;
    \sum_{t\in \{\triangleright,\,\triangleleft\}} B_t^p, \;\;\;\;
    A_s^p \; = \; \frac{1-A_s}{2}\frac{1+B_\triangleright}{2}\frac{1+B_\triangleleft}{2},
    \;\;\;\;
    B_t^p \; = \; \frac{1-B_t}{2}.
\end{equation}
Since all terms in the Hamiltonian commute with one another, they can be simultaneously measured. If an $m$-anyon is present on a star operator $A_s^p$, i.e., $B_\triangleright^p = 1$ or $B_\triangleleft^p = 1$, the measurement of that star yields a value of 0. This commuting-projector Hamiltonian formulation is particularly useful for decoding in the presence of measurement errors, but the operator $A_s^p$ is more difficult to measure experimentally than $A_s$.\\
\indent In our numerical simulations, we adopt the quasistabilizer Hamiltonian formulation of the $D_4$ TO.
\subsection{Constraints on $e$-charge measurements following single color of Pauli $\hat{X}$ errors}
\indent In this subsection, a simplified version of the stabilizer formalism for the $D_4$ TO developed in Ref.~\onlinecite{chirame2024stabilizingnonabeliantopologicalorder} is used to derive the parity constraints on the intermediate $e$-charge measurements after the introduction of red Pauli $\hat{X}$ errors. Under this noise channel, blue and green star operators are affected, as well as the red triangle operators located on those stars. The blue and green stars together form a honeycomb lattice, with each color occupying one of its two sublattices.\\
\indent The ground states of a $D_4$ TO do not have anyon excitations. Therefore, the stabilizer generators of the ground states on a torus consist of the Hamiltonian terms, namely the star and triangle operators in Eq.~\eqref{Hamiltonian_terms}, along with six logical operators \cite{Yoshida, Iqbal2024}. After the application of a short red Pauli $\hat{X}$ string, the stabilizers are modified as follows:
\begin{equation}
    \vcenter{\hbox{\includegraphics[height=10em]{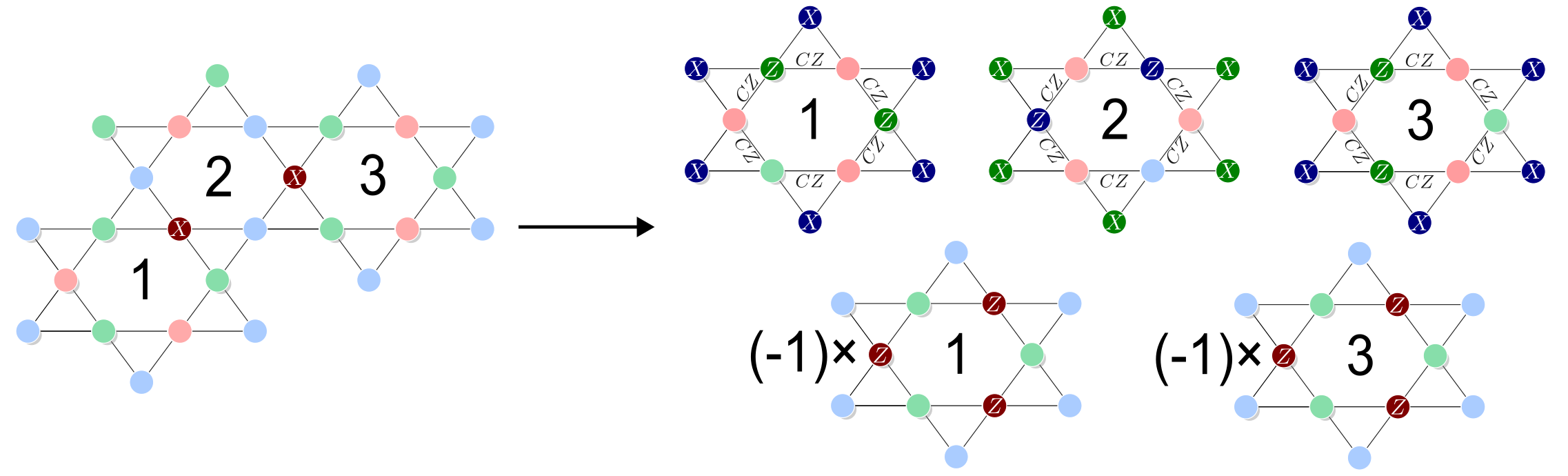}}},
    \label{after_error}
\end{equation}
while all other stabilizers are unchanged. Since the Pauli $\hat{X}$ string ends on stars 1 and 3, these two stars will not be measured. The stabilizer $p_2A_2$ resulting from the measurement of star 2 anti-commutes with the modified star stabilizers 1 and 3. Consequently, the product of the modified star stabilizers 1 and 3, together with $p_2A_2$, becomes the new stabilizer generators following the measurement, replacing modified star stabilizers 1 and 3:
\begin{equation}
    \vcenter{\hbox{\includegraphics[height=6em]{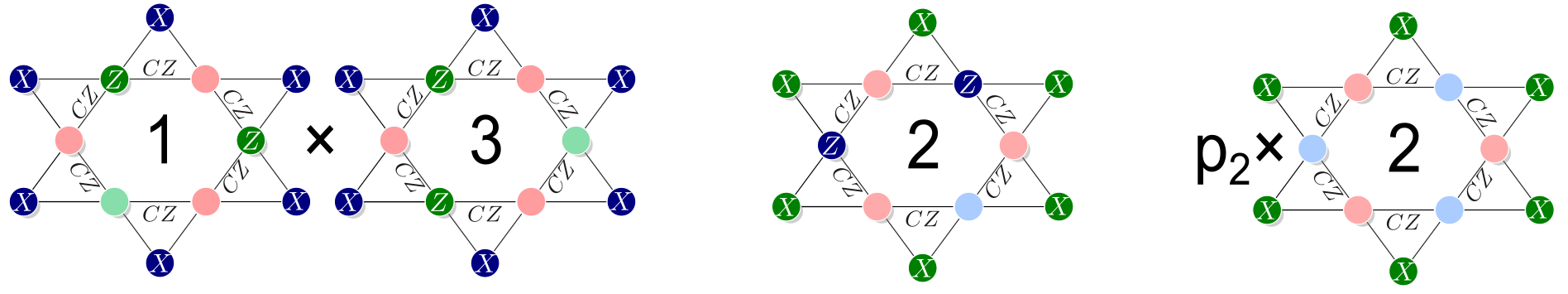}}},
    \label{after_measurement}
\end{equation}
where the measurement outcome, $p_2$, is $+1$ (vacuum) or $-1$ ($e$-charge) with equal probability $\frac{1}{2}$. The triangle stabilizers commute with the measurement of star 2 and are therefore unaffected.\\
\indent A parity constraint is relevant when the Pauli $\hat{X}$ string forms a closed loop. Following the analysis above, the star stabilizers are replaced by the following stabilizers after the measurement of stars 2 and 4:
\begin{equation}
    \vcenter{\hbox{\includegraphics[height=10.5em]{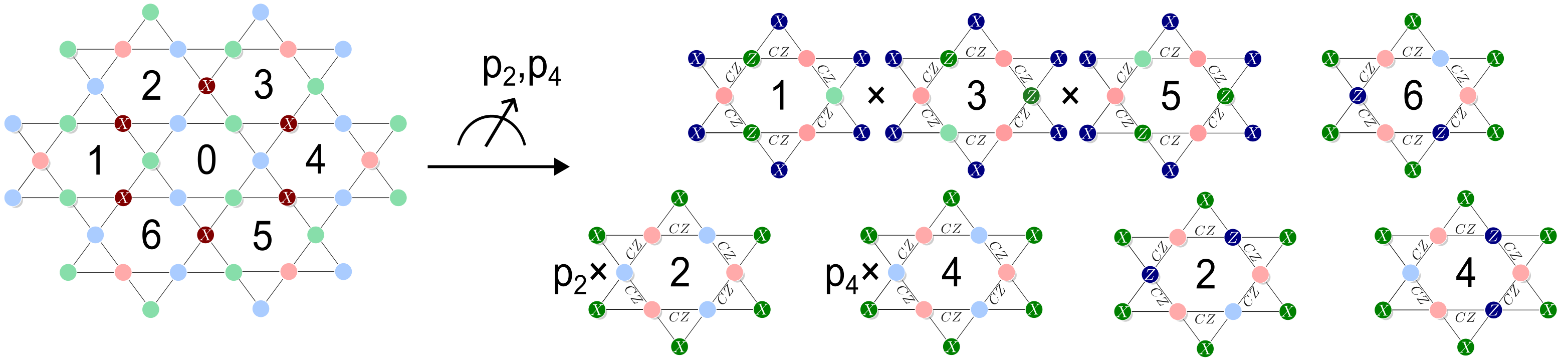}}},
    \label{loop_after_meas}
\end{equation}
while the triangle stabilizers remain unchanged. Since star 6 commutes with all stabilizers in Eq.~\eqref{loop_after_meas}, it must be a product of stabilizers up to a constant:
\begin{equation}
    \vcenter{\hbox{\includegraphics[height=4.2em]{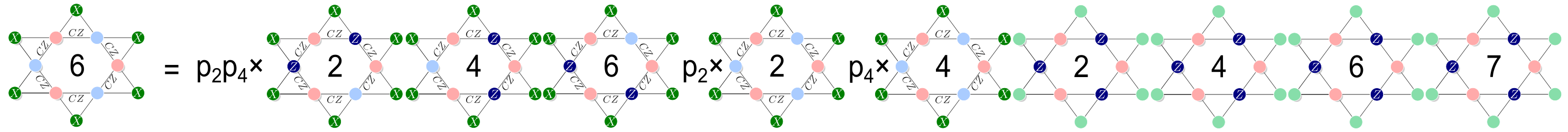}}}.
\end{equation}
This leads to a deterministic outcome $p_6 = p_2p_4$, reflecting the requirement that the total number of green $e$ charges on stars 2, 4, and 6 is even. A similar analysis shows that the total number of blue $e$-anyons on stars 1, 3, and 5 must also be even.\\
\indent One can also consider the case where the error string forms a branched loop:
\begin{equation} \label{eq:kagome}
    \vcenter{\hbox{\includegraphics[height=9em]{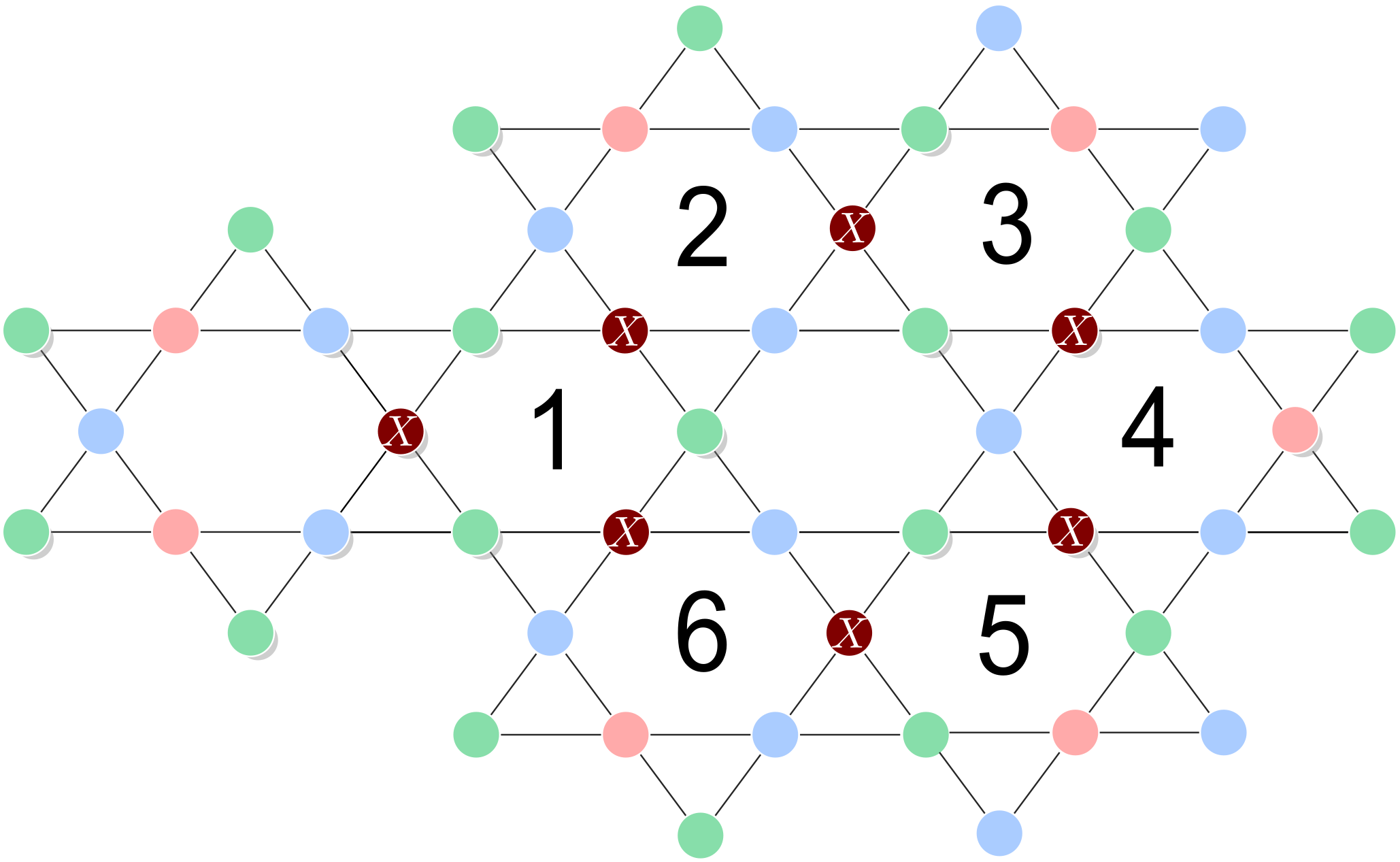}}}.
\end{equation}
Using the stabilizer formalism, one can show that the even-parity constraint on green stars 2, 4, and 6 still holds. However, since star 1 supports an $m$-anyon and is therefore not measured, the constraint on blue stars 1, 3, and 5 no longer applies, and the measurement outcomes of stars 3 and 5 can take arbitrary values of $\pm1$.\\
\indent Therefore, there is one even-parity constraint for each non-branching, homologically trivial Pauli $\hat{X}$ loop for each color of $e$-anyons. This is the only constraint on the measurement outcomes after the introduction of Pauli $\hat{X}$ errors.\\
\indent On a non-branching Pauli $\hat{X}$ loop with nontrivial homology, the constraints on $e$-anyon measurement depend on the logical sector of the initial state that we are trying to protect. One might suspect that this would lead to different error-correction thresholds for different logical sectors. However, an unbranched homologically nontrivial loop is extremely unlikely, and such a loop is not observed in any of the over $10^8$ error configurations in our simulation of the matching decoder. Therefore, it was concluded that all ground states of the $D_4$ TO have the same error-correction threshold, and we did not specify the initial state on which error correction was performed in this paper.
\subsection{\texorpdfstring{Value of $P(\bm{s}|E)$ in Eq.~\eqref{Probability}}{Value of P(E|s) in Eq.~\eqref{Probability}}}
\indent For the $D_4$ TO, Eq.~\eqref{Probability} in the main text can be rewritten as
\begin{equation}
    P(\bm{s}|E)=\bra{E} \prod_{s\in \{\textrm{\ding{65}}\}} \left[ (1-\lambda_s)(1-A_s^p) + \lambda_sA_s^p \right] \prod_{t\in \{\triangleright,\,\triangleleft\}} \left[ (1-\lambda_t)(1-B_t^p) + \lambda_tB_t^p \right] \ket{E},
    \label{D4Probability}
\end{equation}
where $\bm{s}=\{\bm{m}_{R}, \bm{e}_{B}, \bm{e}_{G}\}$ under red Pauli $\hat{X}$ errors. The state $\ket{E}$ is an eigenstate of the triangle operators, as shown in Eq.~\eqref{after_error}, with $\lambda_{t\in\partial E}=1$ at the endpoints of the error string and $\lambda_{t\notin\partial E}=0$ elsewhere, corresponding to syndromes $\bm{m}_{R}=\partial E$. Furthermore, projectors $A_s^p$ have an eigenvalue $\lambda_s=0$ at these endpoints, $s\in\partial E$, and away from the error string, $s\notin E$.\\
\indent Whenever the state is in an eigenstate of a commuting projector $A_s^p$ or $B_t^p$, the expectation values of $(1-\lambda_s)(1-A_s^p) + \lambda_sA_s^p$ or $(1-\lambda_t)(1-B_t^p) + \lambda_tB_t^p$ are equal to one. Therefore,
\begin{equation}
    P(\bm{s}|E)=\bra{E} \prod_{s\in \{\bivalent\}=E\setminus \partial E} \left[ (1-\lambda_s)(1-A_s^p) + \lambda_sA_s^p \right] \ket{E},
    \label{D4Probability2}
\end{equation}
where ${\bivalent}$ denotes the set of blue and green stars on which Pauli $\hat{X}$ errors have occurred on exactly two red qubits, and the relation $\{\bivalent\}=E\setminus \partial E$ holds due to the geometry of the kagome lattice.\\
\indent On each star $s \in \bivalent$, a measurement of $A_s^p$ is performed, producing the results $0$ (vacuum) and $1$ ($e$-charge), each with equal probability $\frac{1}{2}$. Equivalently, this corresponds to the operator $(1 - \lambda_s)(1 - A_s^p) + \lambda_s A_s^p$ having expectation value $\frac{1}{2}$ for both $\lambda_s = 0$ and $\lambda_s = 1$. However, when the error string $E$ forms an isolated closed loop of any color, the $e$-charge measurement outcomes are subject to a parity constraint, as demonstrated in the previous subsection. When the constraint is violated by the syndromes $\bm{s}$, the expectation value in Eq.~\eqref{D4Probability2} vanishes, indicating that $E$ is not a valid error string to be considered for optimal decoding. When the syndromes satisfy the constraint, the measurement outcome of one star along the isolated closed loop is determined by the outcomes of all other stars on the loop. Consequently, the expectation value of $(1 - \lambda_s)(1 - A_s^p) + \lambda_s A_s^p$ for this particular star is one.\\
\indent Denoting the total number of parity constraints in $E$ as $C$, we can determine the value of $P(\bm{s}|E)$ by
\begin{equation}
    P(\bm{s}|E)=2^{C-\bivalent}=2^{C+\trivalent+\frac{|\bm{m}|}{2}-|E|},
    \label{D4probability3}
\end{equation}
where $\trivalent$ is the number of Y-shaped intersections in $E$ which always host an $m$-flux, $\bm{m}$ is the total number of red $m$-fluxes in the syndrome $\bm{s}$, and $|E|$ is the length of the error string. Eq.~\eqref{D4probability3} is valid as long as all constraints $C$ are satisfied by the syndromes $\bm{s}$ and $\bm{m}_{R}=\partial E$; otherwise, $P(\bm{s}|E) = 0$, thereby disallowing the error string $E$.\\
\indent In our numerical simulations where only red Pauli $\hat{X}$ errors are considered, we use Eq.~\eqref{D4probability3}, whereas the corresponding expression for Eq.~\eqref{Probability} in the main text under the full Pauli error channel, along with its mapping to a local statistical mechanics model, is provided in Appendix~\ref{D4localstatmech}.

\subsection{Constraints on $e$-charge measurements after correcting $m$-fluxes}
\indent In this subsection, we derive the parity constraints on $e$-charge measurements following the correction of $m$-fluxes using red Pauli $\hat{X}$ strings, and discuss their implications for determining logical errors.\\
\indent After the measurements yielding the stabilizers in Eq.~\eqref{after_measurement}, error correction can be performed by applying Pauli $\hat{X}$ operators to the same qubits as the error qubits in Eq.~\eqref{after_error}, namely the red qubits between star 1 and star 2, and between star 2 and star 3. By removing the $m$-fluxes, this operation restores the quantum state to the $+1$ eigenstate of all triangle operators, leading to new stabilizers:
\begin{equation}
    \vcenter{\hbox{\includegraphics[height=5em]{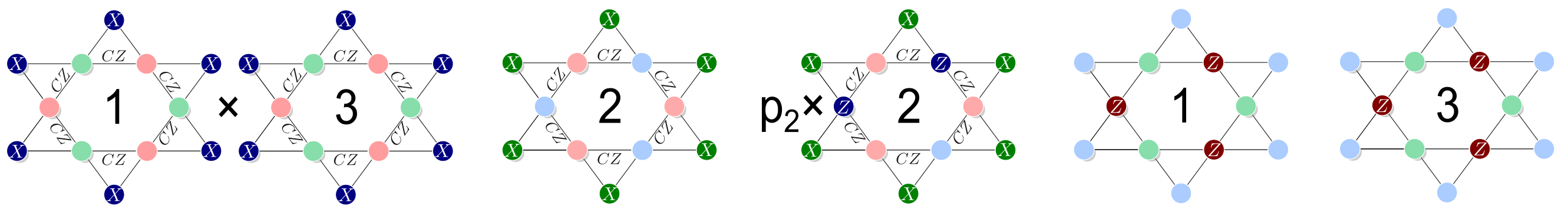}}}.
    \label{after_error_correction}
\end{equation}
All other stabilizers remain unchanged. In the subsequent round of $e$-charge measurements, all star operators commute and can be measured simultaneously, as there are no remaining $m$-fluxes. In particular, the first stabilizer in Eq.~\eqref{after_error_correction} imposes an even-parity constraint on stars 1 and 3, resulting in the same measurement outcome $p$ for both. Furthermore, the stabilizer resulting from the measurement of either star 1 or 3 anti-commutes with only the third stabilizer in Eq.~\eqref{after_error_correction}, and therefore replaces it, resulting in updated stabilizers:
\begin{equation}
    \vcenter{\hbox{\includegraphics[height=5em]{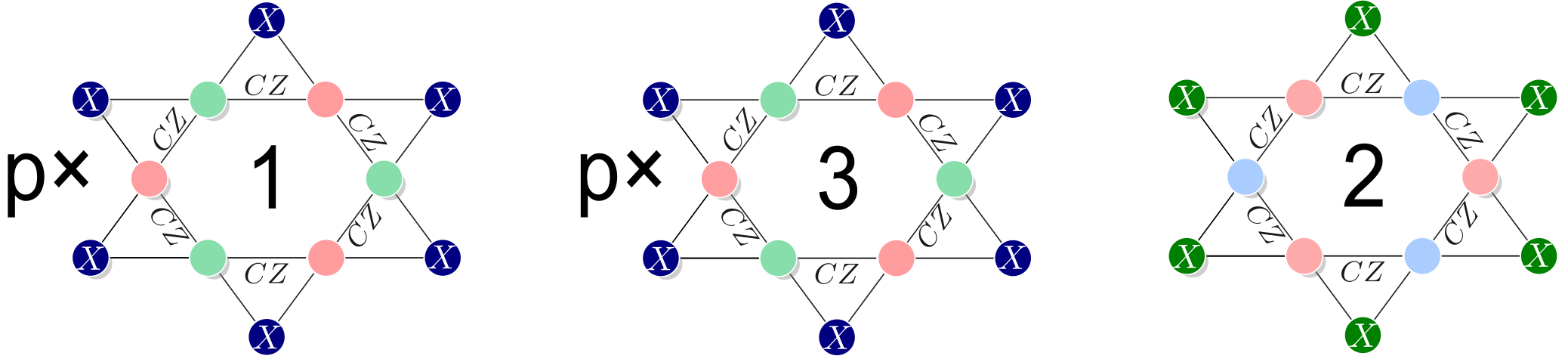}}}.
    \label{final_stabilizer}
\end{equation}
\indent For each connected component in the union of the physical error and correction Pauli $\hat{X}$ strings, two parity constraints on the $e$-charge measurements after flux correction arise, one for the blue stars and one for the green stars. This occurs because the stabilizer resulting from an $e$-charge measurement, whether post-error or post-flux correction, anti-commutes with neighboring star stabilizers linked by the physical error or correction Pauli $\hat{X}$ string, resulting in a new stabilizer given by the product of neighboring stars. The resulting constraints reflect the cumulative effect of two rounds of $e$-charge measurements. Specifically, using the stabilizer formalism described above, it can be shown that, after flux correction, certain sets of star operators become entangled and are governed by the same constraint, according to the following rules:
\begin{center}
\begin{tabular}{ccc}
\hline
Physical error string & Pauli $\hat{X}$ string for & Entanglement of star operators\\
& flux correction & after flux correction\\
\hline
No error & $\topright$ & Entangle neighboring stars to the top and bottom right \\
$\topstring$ & $\bottomright$ & Entangle neighboring stars to the top and bottom right \\
$\bottomleft$ & $\trivalent$ & Entangle neighboring stars to the top and bottom right\\
$\topright$\;, followed by $e$ measurement & No correction & Entangle neighboring stars to the top and bottom right \\
$\topright$\;, followed by $e$ measurement & $\topright$ & Entangle neighboring stars to the top and bottom right\textcolor{red}{*}\\
$\topright$\;, followed by $e$ measurement & $\topleft$ & Entangle all three neighboring stars\\
$\trivalent$ & $\bottomleft$ & Entangle neighboring stars to the top and bottom right\\
\hline
\end{tabular}
\end{center}
These rules apply to each blue or green star operator and its green or blue neighbors, respectively. The case marked with \textcolor{red}{*} corresponds to the one discussed above.\\
\indent When the connected component in the union of the physical error and correction Pauli $\hat{X}$ strings is homologically trivial, the constraints on both blue and green stars are even-parity constraints. In contrast, when the component is homologically nontrivial, the constraints depend on the logical sector of the initial quantum state. Such homologically nontrivial components often signal the occurrence of logical errors.\\
\indent For example, if the initial state is stabilized by the horizontal blue logical $-\bar{Z}$ operator and there exists a single homologically nontrivial horizontal loop formed by the red Pauli $\hat{X}$ physical error and correction strings, then the parity constraint on blue stars along this loop is odd, while that on green stars remains even. The same holds for initial states stabilized by green or vertical logical $-\bar{Z}$ operators. Such a nontrivial loop results in a logical $\bar{X}$ error, which is also manifested in the odd number of $e$ charges remaining after flux correction. In contrast, if the state is stabilized by the vertical blue logical $\bar{X}$ operator and the red Pauli $\hat{X}$ error and correction strings form one or more homologically nontrivial horizontal loops, then no parity constraint is imposed on the blue stars. However, $e$-measurements along any such loop project the system into a definite parity sector of blue $e$ charges, corresponding to a state stabilized by a blue logical $\pm\bar{Z}$ operator, thereby inducing a different type of logical error.\\
\indent Therefore, we declare a logical error whenever the union of the physical error and correction Pauli $\hat{X}$ strings contains any homologically nontrivial component. If no such component is present, we proceed with correcting the remaining $e$ charges.\\
\indent Since no homologically nontrivial component exists in the union of the physical error and correction Pauli $\hat{X}$ strings, there is an even number of $e$ charges of each color. The quantum state after the second round of $e$-measurements is thus equivalent to one generated by any configuration of Pauli $\hat{Z}$ strings that pairwise connect same-color charges along the union, as all such configurations are homologically equivalent. Since the $e$ charges are Abelian, logical $\bar{Z}$ errors arising from homologically nontrivial $e$-charge loops can be identified in the same way as in the toric code. We declare a logical error whenever the symmetric difference between the effective Pauli $\hat{Z}$ error strings and the $e$-charge correction strings forms a noncontractible loop on the torus defined by the periodic boundaries.

\refstepcounter{suppappendix}
\setcounter{equation}{0}
\section{Statistical Mechanics Model for the $D_4$ TO against Single-Qubit Pauli Noise}
\label{D4localstatmech}

\subsection{Pauli $\hat{X}$ noise on red qubits and Pauli $\hat{Z}$ noise on blue and green qubits}
\begin{figure}
    \centering
    \includegraphics[width=0.55\linewidth]{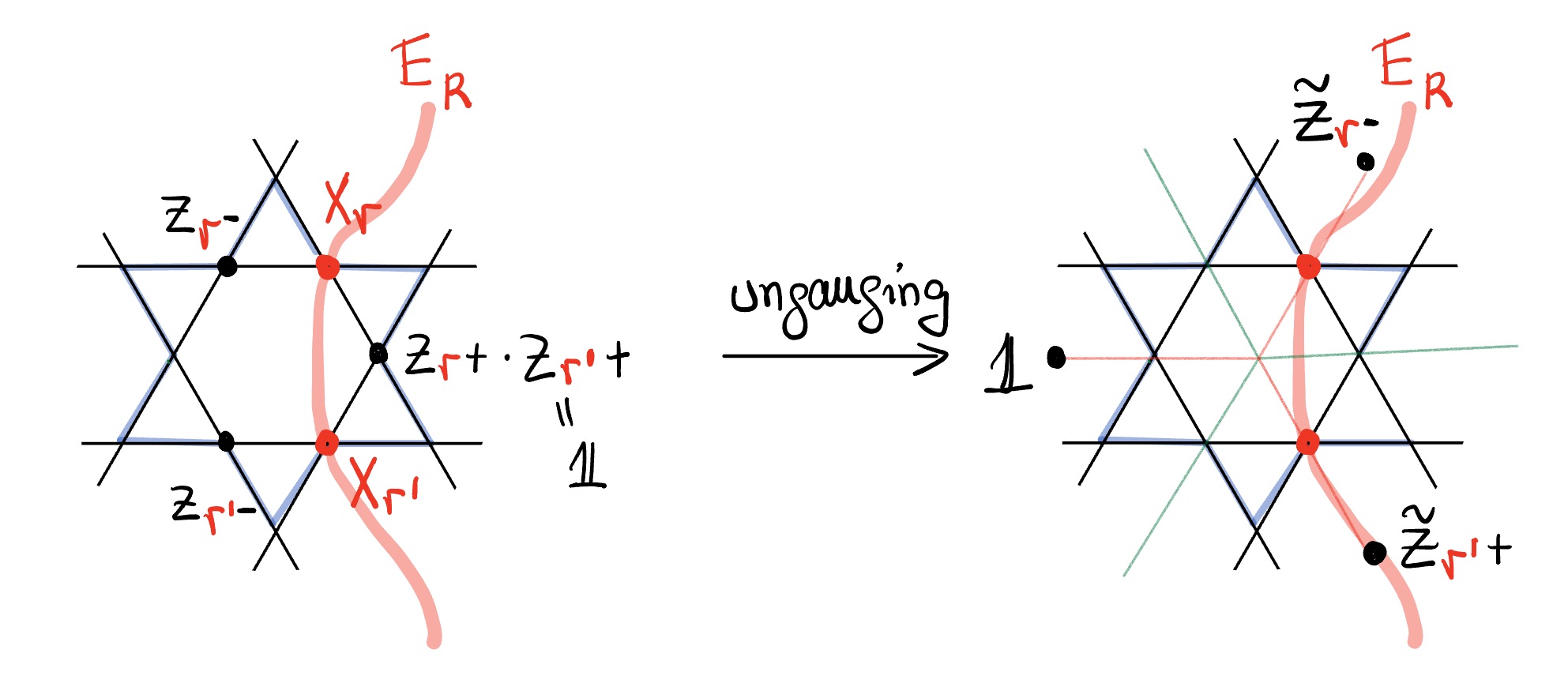}
    \caption{\textbf{Ising strings emerging along $E^{\RR}_X$ errors and ungauging.} The conjugate action of $\prod_{i\in E^{\RR}_X}X_i$ on star operators $A^c_s$ with $c=\GG, \BB$ (as obtained in Eq.~\eqref{neighbor_dressing}) dresses $A^c_s$ with a product of two $Z$ operators. The ungauging then maps these two $\tilde{Z}$ operators lying along the error string $E_X^{\RR}$.}
    \label{fig:ungauging}
\end{figure}

Let us consider the scenario of having Pauli $\R{\hat{X}}$ noise on the \R{red} qubits of the kagome lattice with error rate $p^x_\R{R}$, and Pauli $\hat{Z}$ noise on both \B{blue} and \G{green} qubits with error rates $p^z_\G{G}$ and $p^z_\B{B}$, respectively. The anyon syndromes, denoted by their respective anyon labels, are obtained from measurements of the commuting projectors defined in Eq.~\eqref{Proj_dressing}. We begin by labeling the presence (absence) of an $m_\R{R}$ flux on triangle $t$ by $\R{m_t} = 1(0)$, corresponding to $B_t = -1(1)$ or equivalently $B_t^p = 1(0)$, as defined in Eqs.~\eqref{D4},~\eqref{Hamiltonian_terms}, and~\eqref{Proj_dressing}. The centers of these triangles lie on the vertices of the honeycomb lattice introduced in Appendix~\ref{stabilizer} and represent the mobility of $m_\R{R}$ flux excitations (see upper panels in Fig.~\ref{phase_diagram}). Similarly, we denote the presence or absence of Abelian charges by $e = {\G{e^G_s}, \B{e^B_s}}$, with each variable taking values $1$ or $0$. Our goal in this subsection is to compute the probability of a given error string $E=\{E_X^{\R{R}}, E_Z^{\GG}, E_Z^{\BB} \}$ conditioned on a set of anyon syndromes, and to recast the result as a local statistical mechanics model.\\
\indent To begin, we use that
\begin{equation} \label{eq:1}
    \textrm{prob}(E|\R{m_t}, e) = \textrm{prob}(\R{m_t}, e|E) \textrm{prob}(E)\frac{1}{\textrm{prob}(e,\R{m_t})}.
\end{equation}
Hence, we need to compute $\textrm{prob}(\R{m_t}, e|E) $, since 
\begin{equation}
    \textrm{prob}(E) \propto \left(\underbrace{\frac{p^x_\R{R}}{1-p^x_\R{R}}}_{=t_X^{\RR}}\right)^{E_X^{\RR}} \left(\underbrace{\frac{p^z_\G{G}}{1-p^z_\G{G}}}_{=t_Z^{\GG}}\right)^{E_Z^{\GG}} \left(\underbrace{\frac{p^z_\B{B}}{1-p^z_\B{B}}}_{=t_Z^{\BB}}\right)^{E_Z^{\BB}}.
\end{equation}
\indent To obtain an explicit stat-mech model for the conditional probability, we express $\textrm{prob}(\R{m_t}, e|E)$ as the expectation value of a product of commuting projectors, as given in Eq.~\eqref{Probability}. To detect the presence or absence of a non-Abelian anyon $m_\RR$, we insert the projector $\frac{1}{2}(1 + (1 - 2\R{m_t}) B_t)$ at the center of each red triangle, which is equivalent to $(1 - \R{m_t})(1 - B_t^p) + \R{m_t} B_t^p$. To detect Abelian $e$ anyons, we place the projector $(1 - e_s)(1 - A_s^p) + e_s A_s^p$ at the center of each star $s$. This reduces to $\frac{1}{2}(1 + (1 - 2e_s) A_s)$ for stars with $s \notin \partial E_X^\RR$, and to $1 - e_s$ for stars with $s \in \partial E_X^\RR$ (and hence vanishes if an $m_\RR$ and an $e$ anyon lie on the same star, namely if $e\in \partial E^{\RR}_X$). Altogether, we find 
\begin{equation} \label{eq:prob_cond_1}
    \textrm{prob}(\R{m_t}, e|E)=\langle E| \prod_{t}\frac{1}{2}(1+(1-2\R{m_t})B_t)\prod_{s\notin \partial E^{\RR}_X}\frac{1}{2}(1+(1-2e_s) {A}_s)|E\rangle,
\end{equation}
where $\ket{E}=\prod_{i\in E^{\RR}_X}X_i \prod_{j\in E^{\GG}_Z \cup E^{\BB}_Z} Z_j|D_4\rangle$, and the condition $s\notin \partial E^{\RR}_X$, can be explicitly imposed by multiplying by $\prod_s(1-\delta_{e_s, \partial E^{\RR}_X})$. To simplify the notation, let us define $\tilde{\R{m_t}}=1-2\R{m_t}\in \{-1,+1\}$ and similarly $\tilde{e}_s=1-2e_s$. Eq.~\eqref{eq:prob_cond_1} then takes the form
\begin{equation}
    \textrm{prob}(\R{m_t}, e|E)=\langle E| \prod_{t}\frac{1}{2}(1+\R{\tilde{m}_t}B_t)\prod_{s\notin \partial E^{\RR}_X}\frac{1}{2}(1+\tilde{e}_s A_s)|E\rangle.
\end{equation}
As a first step, we can simplify the expression by noticing that
\begin{align}
   \nonumber &\textrm{prob}(\R{m_t}, e|E)=\langle D_4|\prod_{i\in E^{\RR}_X}X_i \prod_{j\in E^{\GG}_Z \cup E^{\BB}_Z} Z_j \prod_{t}\frac{1}{2}(1+\R{\tilde{m}_t}B_t)\prod_{s\notin \partial E^{\RR}_X}\frac{1}{2}(1+\tilde{e}_s A_s)\prod_{i\in E^{\RR}_X}X_i \prod_{j\in E^{\GG}_Z \cup E^{\BB}_Z} Z_j|D_4\rangle\\
   \nonumber & = \langle D_4|  \prod_{t\in \partial E^{\RR}_X}\frac{1}{2}(1-\R{\tilde{m}_t}B_t) \prod_{t\notin \partial E^{\RR}_X}\frac{1}{2}(1+\R{\tilde{m}_t}B_t)\prod_{i\in E^{\RR}_X}X_i\prod_{\substack{s\notin \partial E^{\RR}_X\\ s\in \partial E^{\GG}_Z \cup \partial E^{\BB}_Z}}\frac{1}{2}(1-\tilde{e}_s A_s)\prod_{\substack{s\notin \partial E^{\RR}_X\\ s\notin \partial E^{\GG}_Z \cup \partial E^{\BB}_Z}}\frac{1}{2}(1+\tilde{e}_s A_s)\prod_{i\in E^{\RR}_X}X_i |D_4\rangle\\
     &=\prod_{t\in \partial E^{\RR}_X}\frac{1}{2}(1-\R{\tilde{m}_t}) \prod_{t\notin \partial E^{\RR}_X}\frac{1}{2}(1+\R{\tilde{m}_t})\langle D_4|  \prod_{i\in E^{\RR}_X}X_i\prod_{\substack{s\notin \partial E^{\RR}_X\\ s\in \partial E^{\GG}_Z \cup \partial E^{\BB}_Z}}\frac{1}{2}(1-\tilde{e}_s A_s)\prod_{\substack{s\notin \partial E^{\RR}_X\\ s\notin \partial E^{\GG}_Z \cup \partial E^{\BB}_Z}}\frac{1}{2}(1+\tilde{e}_s A_s)\prod_{i\in E^{\RR}_X}X_i |D_4\rangle
\end{align}
where in the last equation we used the fact that $B_t|D_4\rangle = |D_4\rangle$. Here, we explicitly keep the first factors (rather than a Kronecker delta), hoping that we can more easily generalize these expressions to the case of imperfect measurements. We now conjugately apply $\prod_{i\in E^{\RR}_X}X_i$, using the fact that $X_r \textrm{CZ}_{rj} Z_r = \textrm{CZ}_{rj}Z_j$. Note that the error string $E_X^{\RR}$ either does not cross a given star operator $A_s$, since only stars with $s \notin \partial E_X^{\RR}$ contribute to the product, or acts with $\hat{X}$ on two qubits of the same color within the support of $A_s$. Focusing on the latter, we then find
\begin{align} \label{neighbor_dressing}
   \nonumber  &\textrm{prob}(\R{m_t}, e|E)=\prod_{t\in \partial E^{\RR}_X}\frac{1}{2}(1-\R{\tilde{m}_t})(1-\delta_{e_s(t), \partial E^{\RR}_X}) \prod_{t\notin  \partial E^{\RR}_X}\frac{1}{2}(1+\R{\tilde{m}_t})\prod_{\substack{s\notin  E^{\RR}_X\\ s\in \partial E^{\GG}_Z \cup \partial E^{\BB}_Z}}\frac{1}{2}(1-\tilde{e}_s )\prod_{\substack{s\notin  E^{\RR}_X\\ s\notin \partial E^{\GG}_Z \cup \partial E^{\BB}_Z}}\frac{1}{2}(1+\tilde{e}_s )\\
    &\times \langle D_4| \prod_{\substack{s\in E^{\RR}_X  \setminus \partial E^{\RR}_X\\ s\in \partial E^{\GG}_Z \cup \partial E^{\BB}_Z}}\frac{1}{2}\left(1-\tilde{e}_s (\prod_{\R{r}\in s\cup E^{\RR}_X } Z_{\R{r}^+}Z_{\R{r}^-})\right)\prod_{\substack{s\in E^{\RR}_X  \setminus \partial E^{\RR}_X\\ s\notin \partial E^{\GG}_Z \cup \partial E^{\BB}_Z}}\frac{1}{2}\left( 1+\tilde{e}_s (\prod_{\R{r}\in s\cup E^{\RR}_X } Z_{\R{r}^+}Z_{\R{r}^-})\right) |D_4\rangle
\end{align}
where $e_s(t)$ corresponds to the star $s$ on which the triangle $t$ lies, and $\R{r}^{\pm}$ corresponds to the nearest-neighbor sites of a site $\R{r}$ on the central hexagon of the star operator $A_s$ (see Fig.~\ref{fig:ungauging}). The final step is to perform the ungauging map to the $\mathbb{Z}_2^3$ SPT, as introduced in Ref.~\onlinecite{Yoshida} (additional details can be found in Ref.~\onlinecite{Pablo2}), followed by a $\textrm{CCZ}$ disentangling circuit, which maps the topological ground state $|D_4\rangle$ to the trivial product state paramagnet $|+\rangle$ and the product of the two $Z$ operators dressing the star operator, to $\tilde{Z}_{\R{r}^+}\tilde{Z}_{\R{r}^-}$, where $\tilde{Z}_i$ lies on the sites of the honeycomb lattice where the error string $E_X^{\RR}$ lies (see Fig.~\ref{fig:general_statmech}). Overall, we find
\begin{align}
    &\textrm{prob}(\R{m_t}, e|E)=\underbrace{\prod_{t\in \partial E^{\RR}_X}\R{{m}_t}(1-\delta_{e_s(t), \partial E^{\RR}_X})  \prod_{t\notin \partial E^{\RR}_X}(1-\R{{m}_t})}_{=\prod_t\delta_{\R{m_t},\partial E^{\RR}_X}(1-\delta_{e_s(t), \partial E^{\RR}_X}) }\underbrace{\prod_{\substack{s\notin  E^{\RR}_X\\ s\in \partial E^{\GG}_Z \cup \partial E^{\BB}_Z}}{e}_s \prod_{\substack{s\notin  E^{\RR}_X\\ s\notin \partial E^{\GG}_Z \cup \partial E^{\BB}_Z}}(1-{e}_s )}_{=\prod_{s\notin E^{\RR}_X}\delta_{e_s,\partial E^{\GG}_Z \cup \partial E^{\BB}_Z}}\\
    &\times \langle +| \prod_{\substack{s\in E^{\RR}_X  \setminus \partial E^{\RR}_X\\ s\in \partial E^{\GG}_Z \cup \partial E^{\BB}_Z}}\frac{1}{2}\left(1-\tilde{e}_s (\prod_{\R{r}\in s\cup E^{\RR}_X } \tilde{Z}_{\R{r}^+}\tilde{Z}_{\R{r}^-})\right)\prod_{\substack{s\in E^{\RR}_X  \setminus \partial E^{\RR}_X\\ s\notin \partial E^{\GG}_Z \cup \partial E^{\BB}_Z}}\frac{1}{2}\left( 1+\tilde{e}_s (\prod_{\R{r}\in s\cup E^{\RR}_X } \tilde{Z}_{\R{r}^+}\tilde{Z}_{\R{r}^-})\right) |+\rangle,
\end{align}
which can be equivalently rewritten as a positive Boltzmann weight (at zero temperature), using one Ising variable per site
\begin{align}
   \nonumber \textrm{prob}(\R{m_t}, e|E)=&\prod_t\delta_{\R{m_t},\partial E^{\RR}_X}(1-\delta_{e_s(t), \partial E^{\RR}_X}) \prod_{s\notin E^{\RR}_X}\delta_{e_s,\partial E^{\GG}_Z \cup \partial E^{\BB}_Z}\\
    &\times  \frac{1}{2^{|\RR|}}\sum_{\{\sigma\}}\prod_{\substack{s\in E^{\RR}_X  \setminus \partial E^{\RR}_X\\ s\in \partial E^{\GG}_Z \cup \partial E^{\BB}_Z}}\frac{1}{2}\left(1-\tilde{e}_s (\prod_{\R{r}\in s\cup E^{\RR}_X } \sigma_{\R{r}^+}\sigma_{\R{r}^-})\right)\prod_{\substack{s\in E^{\RR}_X  \setminus \partial E^{\RR}_X\\ s\notin \partial E^{\GG}_Z \cup \partial E^{\BB}_Z}}\frac{1}{2}\left( 1+\tilde{e}_s (\prod_{\R{r}\in s\cup E^{\RR}_X } \sigma_{\R{r}^+}\sigma_{\R{r}^-})\right),
\end{align}
where $|\RR|$ denotes the total number of red qubits after applying the ungauging map.

Putting everything together with Eq.~\eqref{eq:1}, and using that $\textrm{prob}(E| \{\R{m_t}, e\}) \textrm{prob}(\{\R{m_t}, e\}) = \textrm{prob}(E\cap \{\R{m_t}, e\})$, we then find
\begin{align} \label{eq:prob_XZZ_noise}
     \nonumber &\textrm{prob}(E\cap \{\R{m_t}, e\}) = (t_X^{\RR})^{E^{\RR}_X} (t_Z^{\GG})^{E^{\GG}_Z} (t_Z^{\BB})^{E^{\BB}_Z}\prod_t\delta_{\R{m_t},\partial E^{\RR}_X} (1-\delta_{e_s(t), \partial E^{\RR}_X})\prod_{s\notin E^{\RR}_X}\delta_{e_s,\partial E^{\GG}_Z \cup \partial E^{\BB}_Z}\\
    &\times \frac{1}{2^{|\RR|}} \sum_{\{\sigma\}} \prod_{\substack{s\in E^{\RR}_X  \setminus \partial E^{\RR}_X\\ s\in \partial E^{\GG}_Z \cup \partial E^{\BB}_Z}}\frac{1}{2}\left(1-\tilde{e}_s (\prod_{\R{r}\in s\cup E^{\RR}_X } \sigma_{\R{r}^+}\sigma_{\R{r}^-})\right)\prod_{\substack{s\in E^{\RR}_X  \setminus \partial E^{\RR}_X\\ s\notin \partial E^{\GG}_Z \cup \partial E^{\BB}_Z}}\frac{1}{2}\left( 1+\tilde{e}_s (\prod_{\R{r}\in s\cup E^{\RR}_X } \sigma_{\R{r}^+}\sigma_{\R{r}^-})\right).
\end{align}

\begin{figure}
    \centering
    \includegraphics[width=0.36\linewidth]{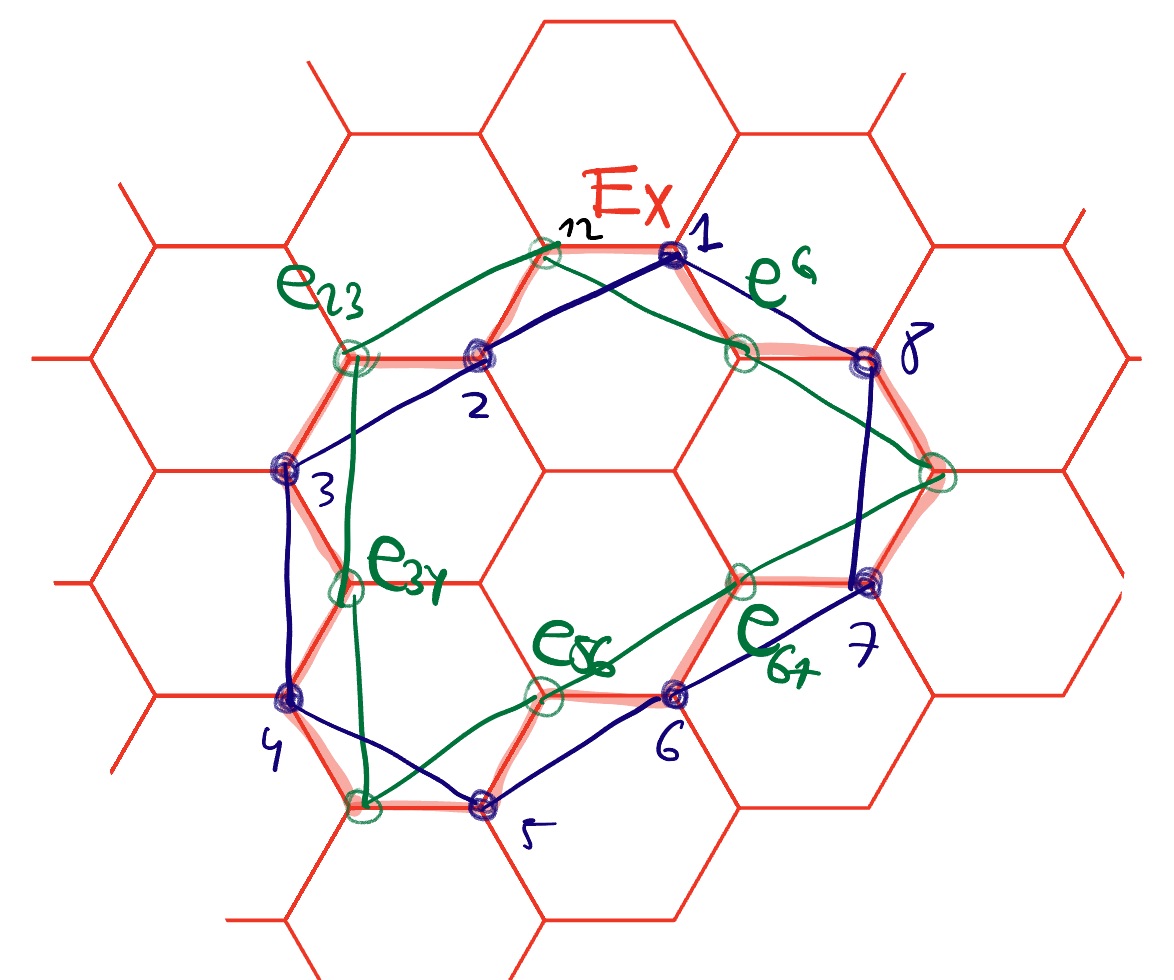}
    \caption{\textbf{$O(4)$ loop model from absence of syndrome fluxes.}}
    \label{fig:simplif}
\end{figure}

From this expression alone, we can directly conclude the following observations:
\begin{itemize}
    \item Non-Abelian $m_\RR$ anyons ($\R{m_t}=1$) need to lie at the boundaries $\partial E^{\RR}_X$.
    \item Abelian $E^{\GG}_Z, E^{\BB}_Z$ anyons ($e_s=1$) need to lie: i) either at the boundaries of $E^{\GG}_Z\cup E^{\BB}_Z $ but not at $E^{\RR}_X$; or ii) $s\in E^{\RR}_X$. 
    \item If we post-select on all $\R{m_t}=0$ for all $t$ (namely $\partial E^{\RR}_X=\emptyset$ and hence all error strings are closed loops $L_\RR$), then assuming no $E^{\GG}_Z, E^{\BB}_Z$ errors, we find that 
    \begin{align}
    & \nonumber \textrm{prob}(L_\RR|\R{m_t}=0\,\forall\,t, e) = (t_X^{\RR})^{L_\RR} \frac{1}{2^{|\RR|}} \sum_{\{\sigma\}} \prod_{\B{s}\in E^{\RR}_X}\frac{1}{2}\left(1+\B{\tilde{e}_s} (\prod_{\R{r}\in \G{s}\cup E^{\RR}_X } \sigma_{\R{r}^+}\sigma_{\R{r}^-})\right)\prod_{\G{s}\in E^{\RR}_X}\frac{1}{2}\left(1+\G{\tilde{e}_s} (\prod_{\R{r}\in s\cup E^{\RR}_X } \sigma_{\R{r}^+}\sigma_{\R{r}^-})\right),
\end{align}
which for closed loops $L_\RR$ takes the value $ \left(\frac{t_X^{\RR}}{2}\right)^{L_\RR} 4^{C_{L_{\RR}}}$, where $C_{L_{\RR}}$ is the number of connected components in $L_\RR$, provided that the parity of anyons along each component is even, i.e., $\prod_{\G{s}\in \ell_\RR}\G{\tilde{e}_s}=\prod_{\B{s}\in \ell_\RR}\B{\tilde{e}_s}=+1$. Such a configuration is represented in Fig.~\ref{fig:simplif}.
\end{itemize}
\subsection{Full Pauli noise on all qubits}
\begin{figure}
    \centering
    \includegraphics[width=0.5\linewidth]{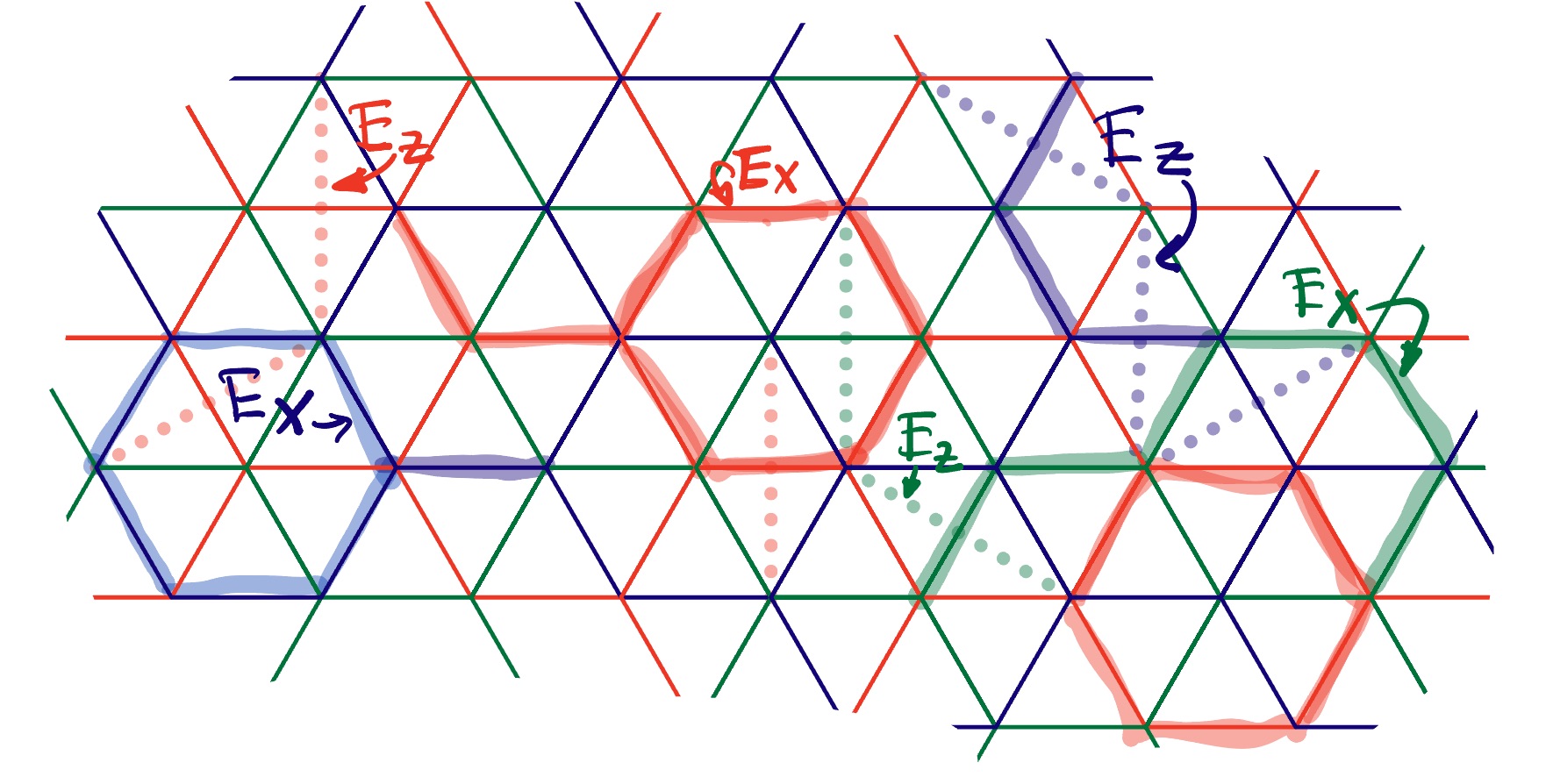}
    \caption{\textbf{Error configuration for general noise model.}}
    \label{fig:general_statmech}
\end{figure}

The previous calculation can be extended to a generic single-site Pauli noise model (with both Pauli $\hat{X}$ and Pauli $\hat{Z}$ channels acting on all sites). Let us denote by $e_s^c, m_t^c$ the charge and flux syndromes associated with a particular color $c =$ \R{$R$}, \G{$G$} or \B{$B$}; and analogously denote by $E^c_X, E^c_Z$ error strings created by Pauli $\hat{X}$ or Pauli $\hat{Z}$ noise, respectively, when acting on qubits of color $c$. We will denote by $m_t, e_s$ the sets of all flux and charge syndromes, respectively, and by $E_X, E_Z$ the sets of all Pauli errors, respectively, across all colors (an example of an allowed error configuration is shown in Fig.~\ref{fig:general_statmech}). Then the conditional probability $\textrm{prob}(\{m_t^c, e_s^c\}|\{E^c_Z, E^c_X\})$ is given by
\begin{align}
  \nonumber & \textrm{prob}(\{m_t^c, e_s^c\}|\{E^c_Z, E^c_X\})=\langle D_4|\prod_{j\in E_X}X_j\prod_{i\in E_Z}Z_i \prod_{t}\frac{1}{2}(1+\tilde{m_t}B_t)\prod_{s\notin \partial E_X}\frac{1}{2}(1+\tilde{e}_s A_s)\prod_{i\in E_Z}Z_i\prod_{j\in E_X}X_j|D_4\rangle\\
   &=\prod_c \delta_{\partial E^c_X, m_t^c}\prod_{c'\neq c}(1-\delta_{e^{c'}_s(t),\partial E_X^c})\langle D_4|\prod_{j\in E_X}X_j\prod_{\substack{s\notin \partial E_X \\ s\in \partial E_Z}}\frac{1}{2}(1-\tilde{e}_s A_s)\prod_{\substack{s\notin \partial E_X \\ s\notin \partial E_Z}}\frac{1}{2}(1+\tilde{e}_s A_s)\prod_{j\in E_X}X_j|D_4\rangle.
\end{align}
Following the same reasoning as in the previous subsection, and denoting by $c, c', c''$ pairwise unequal colors, we find that
\begin{align}
  \nonumber & \textrm{prob}(\{m_t^c, e_s^c\}|\{E^c_Z, E^c_X\})=\prod_{c=\RR,\GG,\BB} \left( \prod_t\delta_{\partial E^c_X, m_t^c}\prod_{c'\neq c}(1-\delta_{e^{c'}_s(t),\partial E_X^c}) \prod_{s\notin E_X^{c'}\cup E_X^{c''}}\delta_{e_s^c,\partial E^{c}_Z }\right)\\
  &\times \langle D_4|\prod_{c=\RR,\GG,\BB}\prod_{\substack{s\in E_X^{c'}\cup E_X^{c''}\\ s\notin \partial E_X^{c'}\cup \partial E_X^{c''} }}\frac{1}{2}\left(1+(-1)^{\delta_{e_s^c,\partial E_Z^c}}e_s^c\prod_{j_{c'}\in s\cap E_X^{c'}}Z_{j_{c'}^+}Z_{j_{c'}^-}\prod_{j_{c''}\in s\cap E_X^{c''}}Z_{j_{c''}^+}Z_{j_{c''}^-}\right)|D_4\rangle,
\end{align}
where $j_{c}^{\pm}$ again correspond to the nearest-neighbor sites of a site $j_c$ of color $c$, lying on the central hexagon of the star operator as for the derivation in Eq.~\eqref{neighbor_dressing}. Hence, $j_c^{\pm}$ lie on qubits of a different color. Notice that unlike in Eq.~\eqref{neighbor_dressing}, now a given star operator $A_s^c$ can be dressed by the nontrivial action of error strings $E_X^{c'}$, $E_X^{c''}$ corresponding to two different colors, when crossing through $A_s^c$. Hence, as a result, one finds the product of four (rather than two) $Z$ operators dressing $A_s^c$. Finally, we further simplify this expression by mapping ($|D_4\rangle \to |+\rangle$) via the ungauging map to the $\mathbb{Z}_2^3$ SPT, which maps the dressing $Z$ operators to the product of four $\tilde{Z}$ operators: two lying on the vertices along the $E_X^{c'}$ error string, and the other two along $E_X^{c''}$. Applying the disentangling unitary $\textrm{CCZ}$, one finds the following expression
\begin{align}
  \nonumber & \textrm{prob}(\{m_t^c, e_s^c\}|\{E^c_Z, E^c_X\})=\prod_{c=\RR,\GG,\BB} \left( \prod_t\delta_{\partial E^c_X, m_t^c}\prod_{c'\neq c}(1-\delta_{e^{c'}_s(t),\partial E_X^c}) \prod_{s\notin E_X^{c'}\cup E_X^{c''}}\delta_{e_s^c,\partial E^{c}_Z }\right)\\
  &\times \langle +|\prod_{c=\RR,\GG,\BB}\prod_{\substack{s\in E_X^{c'}\cup E_X^{c''}\\ s\notin \partial E_X^{c'}\cup \partial E_X^{c''} }}\frac{1}{2}\left(1+(-1)^{\delta_{e_s^c,\partial E_Z^c}}e_s^c\prod_{j_{c'}\in s\cap E_X^{c'}}\tilde{Z}_{j_{c'}^+}\tilde{Z}_{j_{c'}^-}\prod_{j_{c''}\in s\cap E_X^{c''}}\tilde{Z}_{j_{c''}^+}\tilde{Z}_{j_{c''}^-}\right)|+\rangle,
\end{align}
where $\tilde{Z}_i$ are defined analogously to the previous section (see Fig.~\ref{fig:ungauging}) but now lying on the triangular lattice displayed in Fig.~\ref{fig:general_statmech}. Alternatively, we can introduce an Ising variable $\sigma=\pm 1$ per vertex on this lattice, and obtain
\begin{align}
  \nonumber & \textrm{prob}(\{m_t^c, e_s^c\}|\{E^c_Z, E^c_X\})=\prod_{c=\RR,\GG,\BB} \left( \prod_t\delta_{\partial E^c_X, m_t^c}\prod_{c'\neq c}(1-\delta_{e^{c'}_s(t),\partial E_X^c}) \prod_{s\notin E_X^{c'}\cup E_X^{c''}}\delta_{e_s^c,\partial E^{c}_Z }\right)\\
  &\times \frac{1}{2^{N}} \sum_{\{\sigma\}}\prod_{c=\RR,\GG,\BB}\prod_{\substack{s\in E_X^{c'}\cup E_X^{c''}\\ s\notin \partial E_X^{c'}\cup \partial E_X^{c''} }}\frac{1}{2}\left(1+(-1)^{\delta_{e_s^c,\partial E_Z^c}}e_s^c\prod_{j_{c'}\in s\cap E_X^{c'}}\sigma_{j_{c'}^+}\sigma_{j_{c'}^-}\prod_{j_{c''}\in s\cap E_X^{c''}}\sigma_{j_{c''}^+}\sigma_{j_{c''}^-}\right),
\end{align}
where $N$ is the number of vertices in the resulting triangular lattice. From this expression one can then obtain $\textrm{prob}(\{E^c_Z, E^c_X\}|\{m_t^c, e_s^c\})$ using Bayes' theorem. 
\refstepcounter{suppappendix}
\setcounter{equation}{0}
\section{Numerical Simulation Details}
\label{Simulation}
\indent In this subsection, we detail the numerical simulations used to determine the error-correction threshold of the $D_4$ TO under Pauli $\hat{X}$ noise on red qubits and Pauli $\hat{Z}$ noise on blue and green qubits. The $D_4$ topological order is defined on an $L \times L$ lattice, where $L$ denotes the number of unit cells along each direction, and each unit cell contains 9 stars, as indicated by the boxed region:
\begin{equation}
    \vcenter{\hbox{\includegraphics[height=10em]{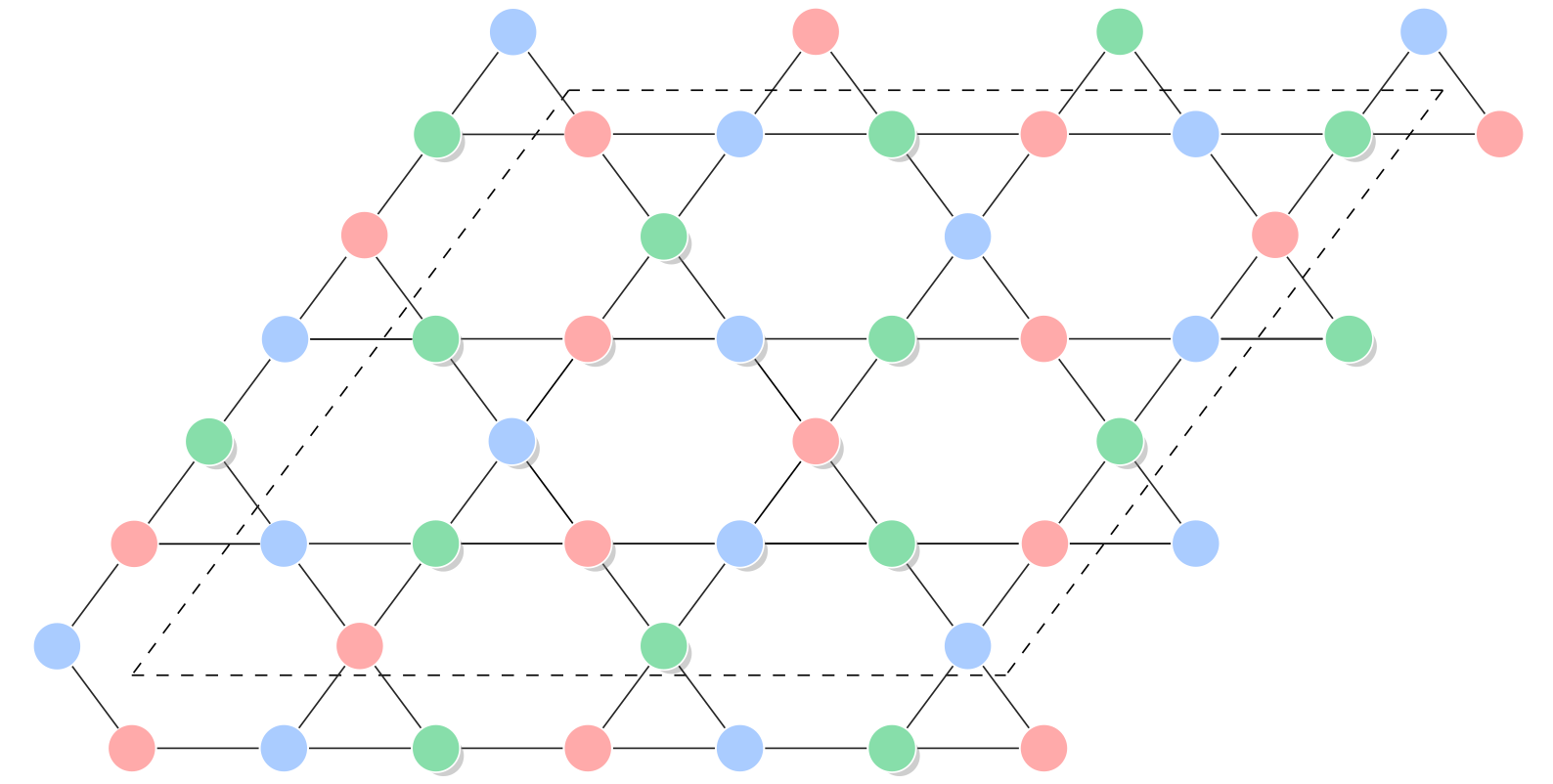}}}
    \label{unit_cell}
\end{equation}
\indent As discussed later, we introduce classical Ising spins at the centers of the hexagons of the honeycomb lattice, whose vertices correspond to blue and green stars, as described in Appendix~\ref{stabilizer} and shown in Fig.~\ref{fig:simplif}, or equivalently, at the vertices of the dual triangular lattice, in order to locally deform error strings.
\subsection{Statistical mechanics models}
\indent In the first three subsections, we consider red Pauli $\hat{X}$ noise, which pair-creates non-Abelian red $m$-fluxes, $\bm{m}_{red}$, at the endpoints of error strings $E$ and leaves a superposition of vacuum and $e$ charges along their paths.\\
\indent Following the seminal work by Dennis \textit{et al.} in Ref.~\onlinecite{Dennis}, the error-correction problem using the unheralded decoders for the $m$-fluxes is typically analyzed by mapping it to a classical random-bond Ising model (RBIM) defined on the dual triangular lattice:
\begin{equation}
    H_{1}(\{\sigma\}) = - \sum_{\langle i,j\rangle} \eta_{ij}(E)\sigma_i\sigma_j, \;\;\;\; \mathcal{Z}_{\bm{m}_R}=\sum_{\{\sigma\}}e^{-\beta H_1},
    \label{RBIM}
\end{equation}
where $\eta_{ij} = -1$ along the error string $E$ and $\eta_{ij} = 1$ elsewhere, with fluctuations in the Ising spins $\{\sigma\}$ generating all homologically equivalent strings satisfying $\partial E=\bm{m}_{red}$. The error-correction threshold of the unheralded decoders corresponds to the phase transition of the quenched-disorder stat-mech model obtained after summing over the disorder variable $\bm{m}_{red}$. In particular, the zero-temperature phase transition corresponds to the unheralded MWPM decoder, while the phase transition along the Nishimori line $\beta = \ln\sqrt{\frac{1 - p}{p}}$ corresponds to the maximum-likelihood decoder, where $p$ is the rate of red Pauli $\hat{X}$ errors. If the decoder incorrectly overestimates the error rate as $3p$, the threshold is instead given by the phase transition along $\beta = \ln\sqrt{\frac{1 - 3p}{3p}}$.\\
\indent With intrinsically heralded decoding, the error-correction string is required to pass through all measured $e$ charges, $\bm{e}_{blue}$ and $\bm{e}_{green}$. This problem can be mapped to the following classical local spin model:
\begin{equation}
    H_{2}(\{\sigma\}) = - \sum_{\langle i,j\rangle} \eta_{ij}(E)w_{ij}(\bm{e}_B, \bm{e}_G)\sigma_i\sigma_j, \;\;\;\; w_{ij}(\bm{e}_B, \bm{e}_G) = 1-n_e K, \;\;\;\; \mathcal{Z}_{\bm{s}}=\sum_{\{\sigma\}}e^{-\beta H_2},
    \label{heraldedRBIM}
\end{equation}
where $n_e\in\{0, 1, 2\}$ is the number of $e$ charges connected to the edge, and the value of $\eta_{ij}(E)$ remains the same as above. Here, heralding strength $K$ is a very large positive number that penalizes error strings bypassing any $e$-charge. In our simulations, it is set to $27L^2$, which is three times the total number of edges in the honeycomb lattice. As before, the phase transitions of the quenched-disorder stat-mech model obtained after summing over the disorder variable $\bm{s}$ correspond to the error-correction thresholds of this decoder.\\
\indent To construct the optimal decoder for the $D_4$ TO under red Pauli $\hat{X}$ noise, one must account for the conditional probability $P(\bm{s}|E)$ in Eqs.~\eqref{Probability} and~\eqref{D4probability3}, which leads to
\begin{equation}
    \mathcal{Z}_{\bm{s}}=\sum_{\text{allowed}\;\{\sigma\}}2^{C-\bivalent}e^{-\beta H_2},
    \label{partition}
\end{equation}
where the summation runs over all error strings that satisfy $\partial E=\bm{m}_{red}$ and the parity constraints on $e$ charges. Although the number of constraints $C$ is nonlocal, the factor $2^{\bivalent}$ is local and can be absorbed into the definition of the Hamiltonian, leading to:
\begin{equation}
    H_{3}(\{\sigma\}) = - \sum_{\langle i,j\rangle} \eta_{ij}(E)\tilde{w}_{ij}(\bm{s})\sigma_i\sigma_j, \;\;\;\; \tilde{w}_{ij}(\bm{s}) = 1-n_e K+ \frac{2-n_m}{2} \frac{\ln{2}}{2\beta}, \;\;\;\; \mathcal{Z}_{\bm{s}}=\sum_{\text{allowed}\;\{\sigma\}}2^Ce^{-\beta H_3},
    \label{optimalRBIM}
\end{equation}
where $n_m\in\{0, 1, 2\}$ is the number of $m$-fluxes connected to the edge. The choice of $\frac{\ln{2}}{2\beta}$ is used to implement the factor $2^{-\bivalent}$ such that the partition function in Eq.~\eqref{optimalRBIM} is identical to that in Eq.~\eqref{partition} at all temperatures. The phase transitions of the quenched-disorder stat-mech model away from the Nishimori line correspond to a decoder that incorrectly estimates the error probability $p$ but correctly accounts for the factor $2^{C-\bivalent}$. This is meaningful because the factor arises from the fusion channels of the non-Abelian anyons, which are typically known.\\
\indent We note that in the $p \to 0$ limit, the Hamiltonian $H_3$ in Eq.~\eqref{optimalRBIM} reduces to that of the clean Ising model; however, the $2^C$ factor remains. Since the clean Ising model contains only closed domain walls (as there are no anyons), and each closed loop satisfies exactly two independent constraints (one for each color), the $2^C$ factor contributes a multiplicative factor of $4$ per domain wall. In the $p \to 0$ limit, the model reduces to an $O(4)$ loop model on the honeycomb lattice, which is dual to the triangular lattice on which the Ising spins reside. Since this is known to always be in the short-loop phase \cite{Nienhuis82} (at least for the physically accessible parameter regime), the corresponding model in Eq.~\eqref{optimalRBIM} remains in the ordered phase. We thus conclude that the critical temperature $T_c$ diverges as $p \to 0$.

\subsection{MWPM simulation details}
\indent We first studied the error-correction threshold of the MWPM decoders for correcting red non-Abelian $m$-fluxes arising from red Pauli $\hat{X}$ noise. These thresholds correspond to the zero-temperature phase transition of the stat-mech models in Eqs.~\eqref{RBIM},~\eqref{heraldedRBIM}, or~\eqref{optimalRBIM}. For each MWPM decoder, $N_E$ random error configurations were generated on the edges of an $L\times L$ honeycomb lattice with qubit error rate $p$, which was varied in increments of $0.001$ (or $0.1\%$) around the threshold, with at least four values sampled both above and below it. If necessary, $e$-charge measurements were simulated on star operators that do not host any $m$-flux, with the constraints discussed in Appendix~\ref{stabilizer} enforced using a disjoint-set union data structure and a depth-first search algorithm. The matching of $m$-fluxes was then performed on the honeycomb lattice using the PyMatching Python package \cite{higgott2021pymatchingpythonpackagedecoding}. The edge weights on the lattice were set to $1$ for the unheralded MWPM decoder, to $w_{ij}(\bm{e}_B, \bm{e}_G)$ for the heralded MWPM decoder in Eq.~\eqref{heraldedRBIM}, and to $\tilde{w}_{ij}(\bm{s})$ for the zero-temperature phase transition of the stat-mech model in Eq.~\eqref{optimalRBIM}, where the factor of $2^C$ was ignored in the simulation of the latter, since the MWPM algorithm disfavors isolated closed loops that give rise to the constraints. A logical error was declared if the union of the physical error string and the flux correction string contained any homologically nontrivial component, as determined by a breadth-first search algorithm on the universal cover. Following the finite-size scaling analysis in Ref.~\onlinecite{WANG200331}, the error-correction threshold $p_c$ and the critical exponent $\nu$ were determined by fitting the logical error rates $P_{\text{logical}}$ for each decoder to
\begin{equation}
    P_{\text{logical}} = f(x), \;\;\;\; x=(p-p_c)L^{1/\nu}.
    \label{scaling}
\end{equation}
\indent We also studied the threshold while accounting for logical errors arising from both non-Abelian flux correction and Abelian charge correction. The introduction of physical errors, syndrome measurements, correction using red Pauli $\hat{X}$ strings, and the identification of logical errors due to flux correction were carried out as described above. If no logical error is declared, the simulation proceeds by identifying the blue and green $e$ charges that arise after flux correction using a disjoint-set union data structure, along with the corresponding effective blue and green Pauli $\hat{Z}$ error strings, following the method outlined in Appendix~\ref{stabilizer}. Another round of MWPM decoding was used to determine the shortest Pauli $\hat{Z}$ error-correction string, and a logical error was declared if the symmetric difference between the effective Pauli $\hat{Z}$ error strings and the $e$-charge correction strings contained any noncontractible loop. No logical error is recorded if none is declared after either the flux correction or the charge correction. The threshold $p_c$ and the critical exponent $\nu$ were similarly extracted by fitting to Eq.~\eqref{scaling}.\\
\indent When fitting the numerical data to Eq.~\eqref{scaling}, the function $f(x)=A+Bx+Cx^2+EL^{-1/\mu}$ was first used, following the approach in Ref.~\onlinecite{WANG200331}. This model is considered valid if the fitted value of $\mu$ differs from zero by at least one standard deviation, which is the case when considering both flux and charge correction steps. Otherwise, the distance-independent function $f(x)=A+Bx+Cx^2$ or $f(x)=A+Bx+Cx^2+Dx^3$ was used. For the MWPM decoders, both the quadratic and cubic fits yielded similar residual sums of squares, indicating that the quadratic fit $f(x)=A+Bx+Cx^2$ was sufficient.\\
\indent The parameters and results of the MWPM simulations are summarized in the table below:
\begin{center}
\begin{tabular}{|c|c|c|c|c|c|c|}
\hline
Data point & L & \; $N_E$ \; & $p_c$ & $\nu$ & $\mu$ & Fitting model\\
\hline
Unheralded MWPM & 10-30 even & $10^6$ & 0.15860(1) & 1.496(8) &  & $A+Bx+Cx^2$\\
Heralded MWPM & 10-28 even & $10^6$ & 0.20842(2) & 1.503(10) &  & $A+Bx+Cx^2$\\
Zero-temperature transition of \ref{optimalRBIM} & 10-28 even & $10^6$ & 0.21196(2) & 1.488(8) &  & $A+Bx+Cx^2$\\
\hline
Unheralded MWPM with $e$-correction & 10-28 even & $10^6$ & 0.15850(27) & 1.541(14) & 0.82(83)  & $A+Bx+Cx^2+EL^{-1/\mu}$\\
Heralded MWPM with $e$-correction & 10-26 even & $10^6$ & 0.20844(49) & 1.604(21) & 0.83(53) & $A+Bx+Cx^2+EL^{-1/\mu}$\\
\hline
\end{tabular}
\end{center}
\indent \indent The critical exponents $\nu$ listed in this table are all consistent with literature values, approximately $1.5(1)$, for the RBIM below the Nishimori line \cite{WANG200331, MELCHERT20111828}.\\
\indent The logical error rates and the finite-size scaling fit for the heralded MWPM decoder with flux correction only, corresponding to the yellow symbol in Fig.~\ref{phase_diagram} of the main text, are shown below as an example demonstrating the validity of the fitting model.
\begin{center}
  \includegraphics[height=18em]{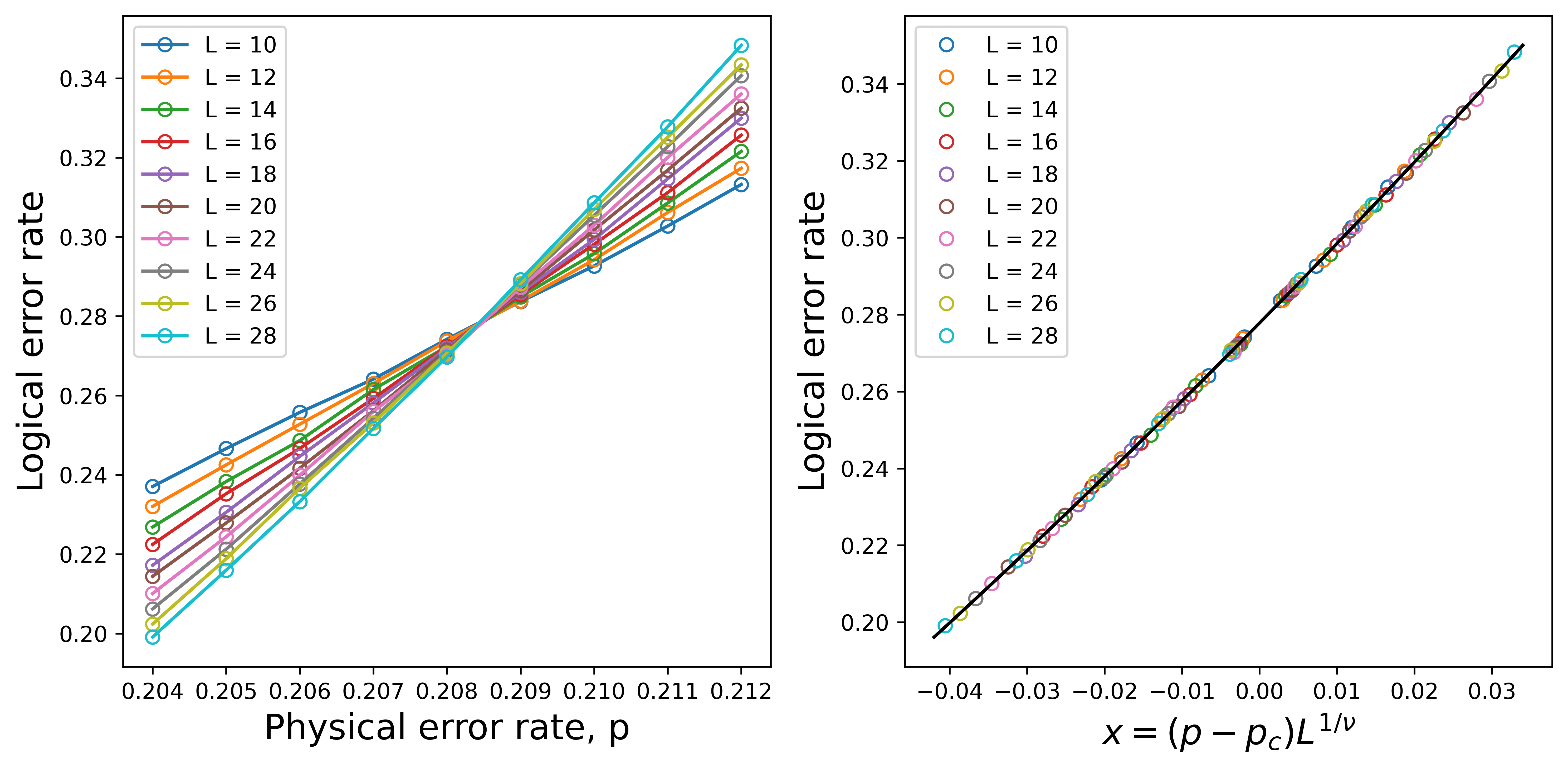}
\end{center}
\subsection{Monte Carlo simulation details}
\indent When all error strings are considered by the decoder, the thresholds for correcting red non-Abelian $m$-fluxes correspond to the finite-temperature phase transitions of the stat-mech models in Eqs.~\eqref{RBIM},~\eqref{heraldedRBIM}, and~\eqref{optimalRBIM}, which were numerically determined using Monte Carlo methods.\\
\indent For each simulation, $N_E$ red Pauli $\hat{X}$ error configurations were generated on the $L\times L$ honeycomb lattice, and the $e$ charges were measured as described in the previous subsection. For each error and syndrome configuration, the edge weights of the Hamiltonians in Eqs.~\eqref{RBIM},~\eqref{heraldedRBIM}, and~\eqref{optimalRBIM} were determined accordingly. $N_{eq}$ Metropolis sweeps were performed to equilibrate the system, followed by $N_{MC}$ Monte Carlo sweeps for data collection. In each Metropolis sweep, $3L^2$ Ising spins $\sigma$, the total number of spins on the lattice, were randomly selected one at a time and flipped with probability $\min\{e^{-\beta\;\delta\epsilon},1\}$, where $\delta\epsilon$ is the change in energy resulting from the proposed spin flip.\\
\indent For simulations of the stat-mech models in Eqs.~\eqref{RBIM} and~\eqref{heraldedRBIM}, the magnetization $m=\sum_i \sigma_i$, along with $m^2$, $m^4$, and $m^8$, was calculated after each Monte Carlo sweep. After averaging over all $N_{MC}$ Monte Carlo sweeps and all $N_E$ error and syndrome configurations, the Binder cumulant
\begin{equation}
    B \; = \; 1 \; - \; \frac{\langle m^4 \rangle}{3 \langle m^2 \rangle^2}
\end{equation}
and its standard deviation were computed from the expectation values $\langle m^2 \rangle$, $\langle m^4 \rangle$, and $\langle m^8 \rangle$.\\
\indent For simulations of the stat-mech model in Eq.~\eqref{optimalRBIM}, the calculation of the nonlocal factor $2^C$ is significantly more computationally expensive than performing local spin flips, making it impractical to update $C$ dynamically during each spin flip. To address this issue, the factor $2^C$ is computed only at the end of each Monte Carlo sweep using a disjoint-set union data structure and a depth-first search algorithm, alongside the calculation of the magnetization $m, \; m^2, \; m^4, \; m^8$. A constraint check is also performed via depth-first search, and the factor $2^C$ is set to zero whenever any constraint is violated, thereby excluding disallowed error strings from contributing to the sampling of the partition function. This approach effectively reweights \cite{FS1, FS2, Landau_Binder_2014} the order parameters according to
\begin{equation}
    \langle \mathcal{O} \rangle_R \; = \; \frac{\langle \mathcal{O}\times 2^C \rangle}{\langle 2^C \rangle}, \;\;\;\; \mathcal{O} = m, \; m^2, \; m^4, \; m^8,
\end{equation}
where $\langle \cdot \rangle$ denotes the average over all samples weighted by $e^{-\beta H_3}$, and $\langle \cdot \rangle_R$ denotes the reweighted average. This reweighting did not compromise the validity of our Monte Carlo sampling, as isolated error loops that give rise to the constraints are rare, and the factor $2^C$ is nearly uniformly distributed. Indeed, in the representative case of Eq.~\eqref{optimalRBIM} along $\beta = \ln\sqrt{\frac{1-p}{p}}$, corresponding to the optimal decoder indicated by the orange star in Fig.~\ref{phase_diagram} of the main text, with $L = 6$ and $p = 0.218$, the magnetization is $\langle m \rangle = 0.78846\pm 0.18397$ before weighting and $\langle m \rangle_R=0.78868\pm 0.18387$, showing negligible difference, while $\langle 2^C \rangle = 1.004\pm 0.142$, indicating that the distribution is fairly sharply peaked around unity. The Binder cumulant is then computed using the reweighted expectation values.\\
\indent The error-correction thresholds $p_c$ and the critical exponents $\nu$ were determined by fitting the Binder cumulant for each decoder to
\begin{equation}
    B = f(x), \;\;\;\; x=(p-p_c)L^{1/\nu}.
\end{equation}
The fitting model $f(x)=A+Bx+Cx^2+EL^{-1/\mu}$ showed no statistical dependence on $L$, as the fitted value of $\mu$ remained within one standard deviation of zero. Therefore, the quadratic fit $f(x)=A+Bx+Cx^2$ was used for simulations along $\beta = \ln\sqrt{\frac{1-p}{p}}$ (the Nishimori line), $\beta = \ln\sqrt{\frac{2-p}{p}}$, and $\beta = \ln\sqrt{\frac{1-2p}{2p}}$, where it was found to be sufficient, and the cubic fit $f(x)=A+Bx+Cx^2+Dx^3$ was used for simulations along $\beta = \ln\sqrt{\frac{1-3p}{3p}}$.\\
\indent The parameters and results of the Monte Carlo simulations are summarized in the table below:
\begin{center}
\begin{tabular}{|c|c|c|c|c|c|c|c|}
\hline
Data point & L & \; $N_E$ \; & $N_{eq}$ & $N_{MC}$ & $p_c$ & $\nu$ & Fitting model\\
\hline
\ref{RBIM} along $\beta = \ln\sqrt{\frac{1-p}{p}}$ & 3-9 & $10^5$ & $3\times10^5$ & $3\times10^5$ & 0.1642(3) & 1.55(20) & $A+Bx+Cx^2$\\
\ref{RBIM} along $\beta = \ln\sqrt{\frac{1-3p}{3p}}$ & \; 3-6 \; & $10^5$ & \; $3\times10^5$ \; & \; $3\times10^5$ \; & 0.1033(1) & 1.19(6) & $A+Bx+Cx^2+Dx^3$\\
\hline
\ref{heraldedRBIM} along $\beta = \ln\sqrt{\frac{1-p}{p}}$ & 3-6 & $10^5$ & $5\times10^5$ & $3\times10^5$ & 0.2044(6) & 1.44(32) & $A+Bx+Cx^2$\\
\ref{heraldedRBIM} along $\beta = \ln\sqrt{\frac{1-3p}{3p}}$ & 3-6 & $10^5$ & $3\times10^5$ & $3\times10^5$ & 0.1058(1) & 1.16(7) & $A+Bx+Cx^2+Dx^3$\\
\ref{heraldedRBIM} along $\beta = \ln\sqrt{\frac{2-p}{p}}$ & 3-6 & $10^5$ & $5\times10^5$ & $3\times10^5$ & 0.2180(10) & 1.57(60) & $A+Bx+Cx^2$\\
\hline
\ref{optimalRBIM} along $\beta = \ln\sqrt{\frac{1-p}{p}}$ & 3-6 & $2\times10^5$ & $5\times10^5$ & $3\times10^5$ & 0.2177(7) & 1.70(46) & $A+Bx+Cx^2$\\
\ref{optimalRBIM} along $\beta = \ln\sqrt{\frac{1-3p}{3p}}$ & 3-6 & \; $2\times10^5$ \; & $5\times10^5$ & $3\times10^5$ & 0.1449(2) & 1.25(9) & $A+Bx+Cx^2+Dx^3$\\
\ref{optimalRBIM} along $\beta = \ln\sqrt{\frac{1-2p}{2p}}$ & 3-6 & \; $2\times10^5$ \; & $5\times10^5$ & $3\times10^5$ & 0.1860(3) & 1.33(15) & $A+Bx+Cx^2$\\
\hline
\end{tabular}
\end{center}
\indent \indent Simulations were not performed for the RBIM in Eq.~\eqref{RBIM} below the Nishimori line, as its phase diagram is already well established in the literature \cite{Nishimori, Honecker, Merz, Lessa, Queiroz09, RBIMtriangular}.\\
\indent Still, at or below the Nishimori line, the fitted values of $\nu$ for Eqs.~\eqref{RBIM} and~\eqref{heraldedRBIM} are consistent with literature values of approximately $1.5(1)$ for the RBIM \cite{Merz, Queiroz09, RBIMtriangular}. Above the Nishimori line, the phase transitions of Eqs.~\eqref{RBIM} and~\eqref{heraldedRBIM} are expected to fall within the Ising universality class, where $\nu = 1$. While our fitted critical exponents suggest that the universality classes above and below the Nishimori line are indeed distinct, the values above the line deviate from $\nu = 1$, likely due to the limited range of lattice sizes $L$. For the stat-mech model in Eq.~\eqref{optimalRBIM}, the universality class at phase transition is not known in literature due to the $2^{C-\bivalent}$ factor and need not agree with that of the RBIM.\\
\indent Simulations of the stat-mech model in Eq.~\eqref{optimalRBIM} below the Nishimori line suffered from poor statistics and fit quality due to long equilibration times. This insufficient equilibration tends to overestimate the transition value of $p$, and the poor fits provided no indication that the transition below the Nishimori line could exceed the optimal threshold on the line. Therefore, we plot the phase boundary below the Nishimori line as a straight line in the full phase diagram:
\begin{center}
  \includegraphics[height=28em]{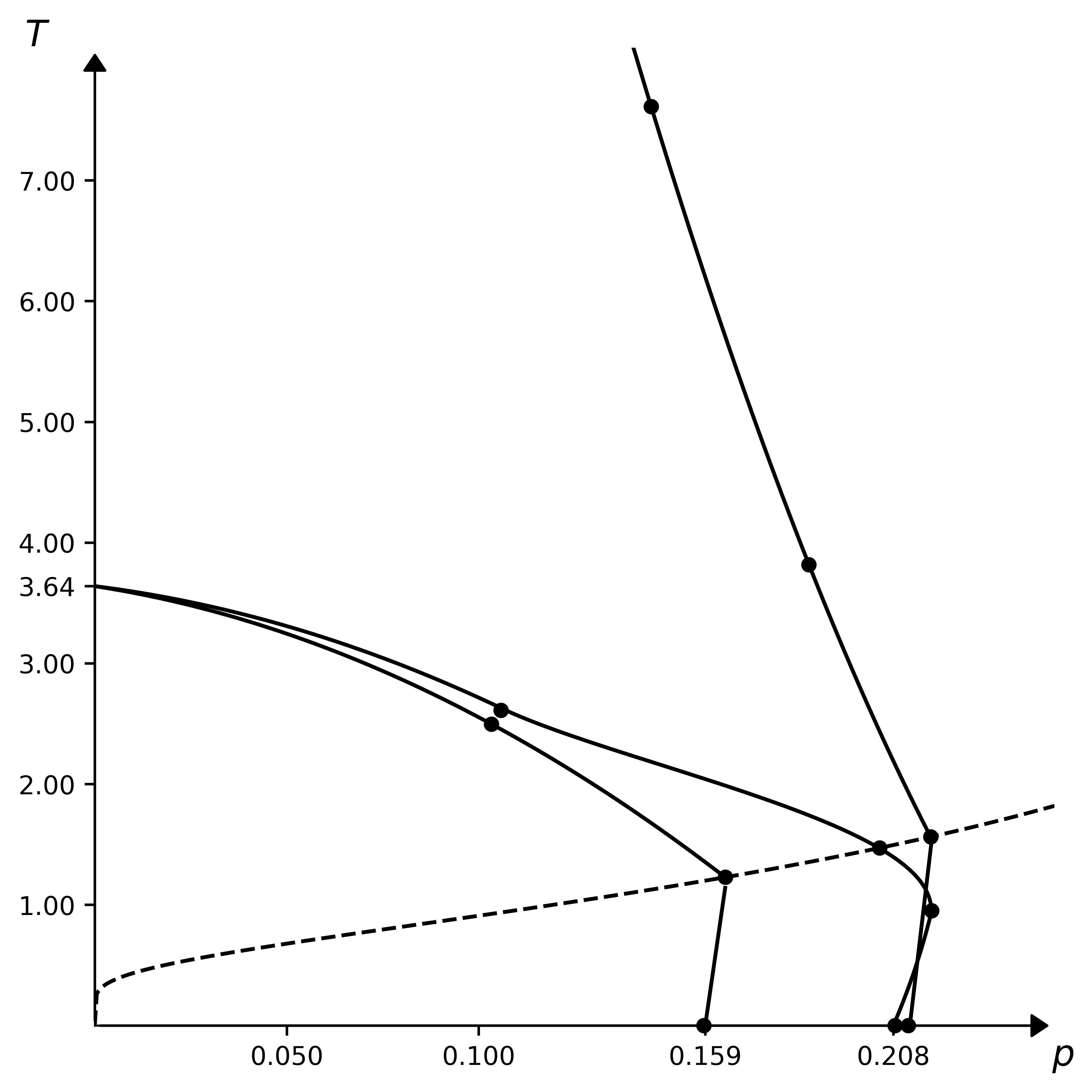}
\end{center}
where the top line corresponds to the phase boundary of the stat-mech model in Eq.~\eqref{optimalRBIM}, the middle line to that in Eq.~\eqref{heraldedRBIM}, the bottom line to that of the RBIM in Eq.~\eqref{RBIM}, and the dashed line is the Nishimori line $\beta = \ln\sqrt{\frac{1-p}{p}}$. Notably, the maximum $p_c$ along the middle phase boundary does not lie on the Nishimori line. This is because the decoders based on Eq.~\eqref{heraldedRBIM} do not account for the probability of $e$-charge measurements and overestimate the likelihood of error string deformation using $\frac{p}{1-p}$ per additional weight, whereas the correct value is $\frac{1}{2}\frac{p}{1-p}$.
\subsection{Decoders against Pauli $\hat{X}$ noise on red qubits and Pauli $\hat{Z}$ noise on blue and green qubits}
\indent Lastly, we consider decoding in the presence of Pauli $\hat{Z}$ noise on blue and green qubits, which pair-creates blue and green $e$ charges and leads to incorrect heralding of the red non-Abelian $m$-fluxes. In this section, we focus exclusively on MWPM decoders, and the reported thresholds on the pair-creation error rate of $m$-fluxes correspond specifically to the correction of red $m$-fluxes.\\
\indent In the physical process, Pauli $\hat{X}$ and Pauli $\hat{Z}$ noise were introduced prior to the measurement of $m$- and $e$-anyon syndromes. This is equivalent to first applying red Pauli $\hat{X}$ errors, then measuring the $m$- and $e$-syndromes, and finally updating the $e$-syndromes to account for the effects of Pauli $\hat{Z}$ errors. Therefore, in each simulation, $N_E$ red Pauli $\hat{X}$ error configurations were generated on the $L\times L$ honeycomb lattice, and the $e$ charges were measured as described previously, before blue and green Pauli $\hat{Z}$ errors were introduced at rate $p_z$, pair-creating blue or green $e$ charges on the end of the Pauli $\hat{Z}$ error strings. If a Pauli $\hat{Z}$ error string terminates at a star hosting an $m$-flux, the syndrome on that star remains unchanged due to the fusion rule $m_R\times e_{B/G}=m_R$.\\
\indent As described in the main text, decoding is performed using an intrinsically heralded MWPM decoder with edge weights $w_{ij}(\bm{e}_B, \bm{e}_G) = 1-n_e K$, where heralding strength $K$ is a tunable parameter. When $K = \infty$, all $e$ charges $\bm{e}_B$ and $\bm{e}_G$ are fully trusted and used to herald the red Pauli $\hat{X}$ error-correction string. For finite positive $K$, such as $K = 0.3$ and $K = 0.6$ used in this work, smaller values of $K$ correspond to reduced confidence in the heralding by the $e$ charges. Logical errors are identified by the presence of homologically nontrivial components in the union of the red Pauli $\hat{X}$ physical error string and the error-correction string, and the corresponding logical error rates are fitted according to Eq.~\eqref{scaling}. The simulation parameters and results for the three choices of heralding strengths $K$ are shown as solid lines in Fig.~\ref{Z_tolerance} of the main text and summarized in the table below.
\begin{center}
\begin{tabular}{|c|c|c|c|c|c|c|}
\hline
Data point & $K$ & L & \; $N_E$ \; & $p_c$ & $\nu$ & Fitting model\\
\hline
$pz=0$ & \; $\infty$ \; & \; 10-28 even \; & \; $10^6$ \; & \; 0.20842(2) \; & \; 1.503(10) \; & $A+Bx+Cx^2$\\
\; $pz=0.0025$ \; & \; $\infty$ \; & 10-24 even & $10^6$ & 0.18740(3) & 1.481(13) & $A+Bx+Cx^2$\\
$pz=0.005$ & \; $\infty$ \; & 10-24 even & $10^6$ & 0.16221(3) & 1.501(17) & $A+Bx+Cx^2$\\
$pz=0.0075$ & \; $\infty$ \; & 10-24 even & $10^6$ & 0.13162(4) & 1.457(21) & $A+Bx+Cx^2$\\
$pz=0.01$ & \; $\infty$ \; & 10-24 even & $10^6$ & 0.09378(7) & 1.522(29) & $A+Bx+Cx^2$\\
\hline
$pz=0$ & \; $0.6$ \; & \; 10-28 even \; & \; $10^6$ \; & \; 0.19372(1) \; & \; 1.489(9) \; & $A+Bx+Cx^2+Dx^3$\\
$pz=0.01$ & \; $0.6$ \; & \; 10-28 even \; & \; $10^6$ \; & \; 0.18408(1) \; & \; 1.481(9) \; & $A+Bx+Cx^2+Dx^3$\\
$pz=0.03$ & \; $0.6$ \; & \; 10-28 even \; & \; $10^6$ \; & \; 0.16437(2) \; & \; 1.501(10) \; & $A+Bx+Cx^2+Dx^3$\\
$pz=0.06$ & \; $0.6$ \; & \; 10-28 even \; & \; $10^6$ \; & \; 0.13305(2) \; & \; 1.494(12) \; & $A+Bx+Cx^2+Dx^3$\\
\hline
$pz=0$ & \; $0.3$ \; & \; 10-28 even \; & \; $10^6$ \; & \; 0.17771(1) \; & \; 1.499(9) \; & $A+Bx+Cx^2+Dx^3$\\
$pz=0.01$ & \; $0.3$ \; & \; 10-28 even \; & \; $10^6$ \; & \; 0.17773(1) \; & \; 1.509(9) \; & $A+Bx+Cx^2+Dx^3$\\
$pz=0.03$ & \; $0.3$ \; & \; 10-28 even \; & \; $10^6$ \; & \; 0.17770(1) \; & \; 1.489(8) \; & $A+Bx+Cx^2+Dx^3$\\
$pz=0.06$ & \; $0.3$ \; & \; 10-28 even \; & \; $10^6$ \; & \; 0.16418(1) \; & \; 1.495(8) \; & \; $A+Bx+Cx^2+Dx^3$ \;\\
\hline
\end{tabular}
\end{center}

\refstepcounter{suppappendix}
\setcounter{equation}{0}
\section{Measurement Errors}
\label{measurement}
\indent We provide an example illustrating how measurement errors in quasistabilizers can be identified from fluctuations of intermediate anyons. This process is trivial for noisy commuting projector measurements. We also discuss the calculation of $P(\bm{s}|E)$ in Eq.~\eqref{Probability} of the main text in the context of continuous error correction, as a step toward constructing an optimal decoder given noisy anyon syndromes.

\subsection{Example of timelike heralding with measurement errors on non-Abelian flux quasistabilizers}
\indent A state within the ground-state subspace of the $D_4$ TO, where quantum information is encoded, is prepared at $t = 0$ with no anyon content. At a later time during the continuous error-correction process, $t = t_0$, Pauli $\hat{X}$ errors happen on four physical qubits
\begin{equation}
    \vcenter{\hbox{\includegraphics[height=7em]{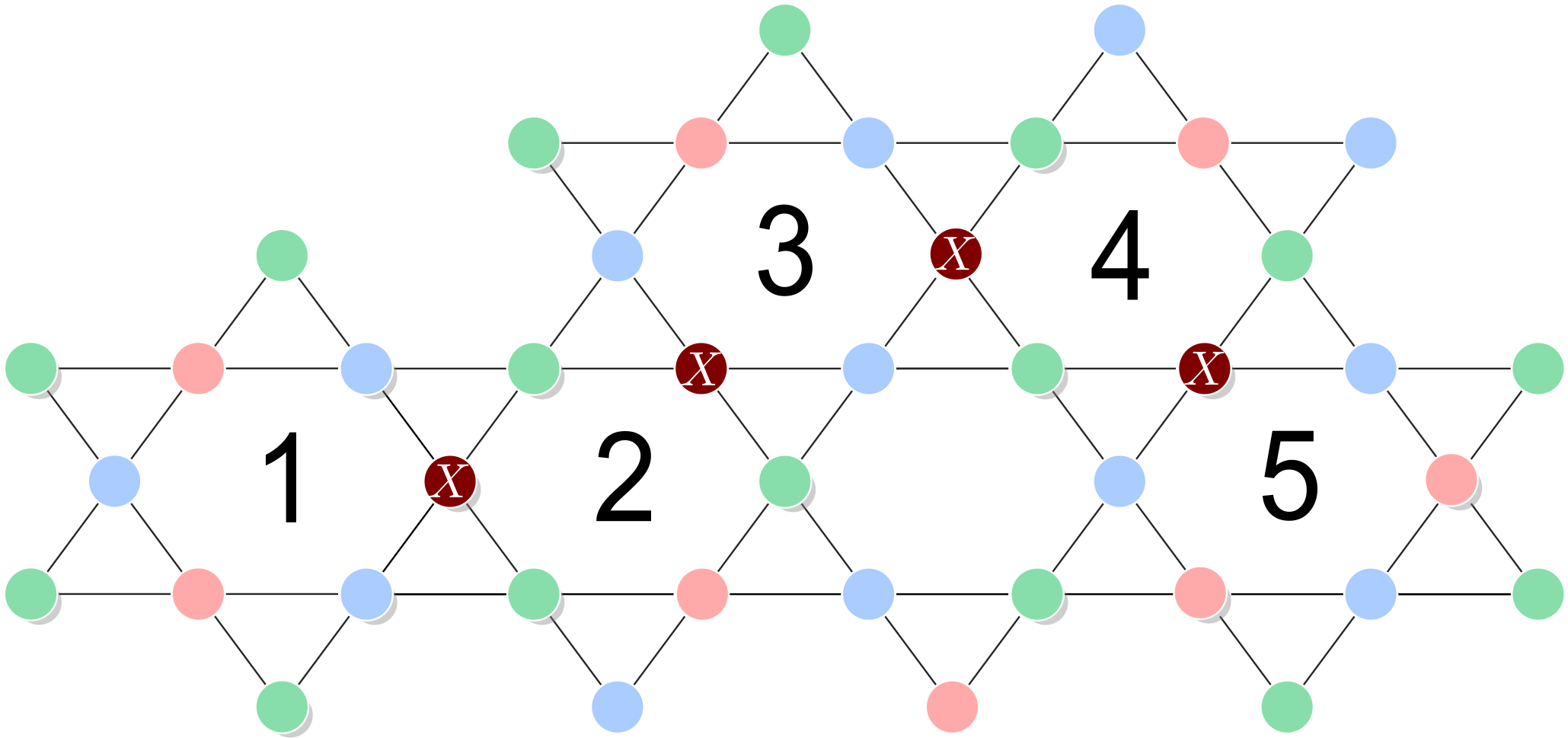}}}
\end{equation}
with the rest of the lattice omitted. Then, at $t = t_0+2$, a false-negative measurement error occurred for the $m$-flux on star 5, resulting in the measurement of the $A_s$ operator on star 5. No other errors occur during the error-correction process from $t = 0$ to $t = t^\prime$. This can result in the following syndromes:
\begin{center}
\begin{tabular}{c|ccccc}
\hline
\;\; Time \;\; & \;\; Star 1 \;\; & \;\; Star 2 \;\; & \;\; Star 3 \;\; & \;\; Star 4 \;\; & \;\; Star 5 \;\; \\
\hline
$t^\prime$ & $m_\triangleright$ & $\mathbf{1}$ & $e$ & $e$ & $m_\triangleright$\\
...... & & & & & \\
$t_0 +4$ & $m_\triangleright$ & $\mathbf{1}$ & $e$ & $e$ & $m_\triangleright$\\
$t_0 +3$ & $m_\triangleright$ & $\mathbf{1}$ & $e$ & $e$ & $m_\triangleright$\\
$t_0 +2$ & $m_\triangleright$ & $\mathbf{1}$ & $e$ & $e$ & $\mathbf{1}$\\
$t_0 +1$ & $m_\triangleright$ & $\mathbf{1}$ & $e$ & $\mathbf{1}$ & $m_\triangleright$ \\
$t_0$ & $m_\triangleright$ & $\mathbf{1}$ & $e$ & $\mathbf{1}$ & $m_\triangleright$ \\
...... & & & & & \\
0 & $\mathbf{1}$ & $\mathbf{1}$ & $\mathbf{1}$ & $\mathbf{1}$ & $\mathbf{1}$ \\
\hline
\end{tabular}
\end{center}
The anyon fluctuations on star 4 and the vacuum measured on star 5 at $t = t_0+2$ can be interpreted as a non-Abelian measurement error on star 5, occurring with probability $q_m$. While this syndrome pattern could also arise from a combination of a non-Abelian measurement error on star 5 and an $e$-charge pair-creation error between star 4 and an omitted star at time $t = t_0+2$, we disregard this possibility, as it is obviously less likely than a single non-Abelian measurement error. Alternatively, aside from a single non-Abelian measurement error, the most probable error patterns capable of producing this syndrome involve at least three errors. One such possibility consists of three $m$-flux pair-creation errors: one between stars 4 and 5 at time $t = t_0+2$, one between star 4 and an omitted star at time $t = t_0+2$, and one between star 5 and an omitted star at time $t = t_0+3$, occurring with probability $p_m^3$. Another possibility involves two $m$-flux pair-creation errors, between star 5 and an omitted star at time $t = t_0+2$ and $t = t_0+3$, along with an $e$-charge pair-creation error between star 4 and an omitted star at time $t = t_0+2$. This occurs with probability $p_m^2p_e$. Typically, the pair-creation error rates are comparable to or smaller than the measurement error rate, i.e., $p_e, p_m \lesssim q_m$. Therefore, errors in the quasistabilizer measurements of non-Abelian anyons can be reliably identified with the help of timelike heralding provided by intermediate anyon fluctuations. In contrast, a measurement error in one commuting projector does not affect the outcomes of other projectors, and thus, there will be no timelike heralding.\\
\indent The identified measurement errors, along with physical errors indicated by syndrome changes at specific time steps, are placed on a three-dimensional (3D) lattice and serve as terminals for the error-correction strings. Error correction then proceeds using 3D decoders. Under pair-creation and measurement errors of non-Abelian fluxes with rates $p$ and $q$, respectively, the spacelike edge weights in the 3D matching decoder remain unchanged from the perfect measurement case, while timelike edges acquire weight $\frac{\ln{\frac{q}{1-q}}}{\ln{\frac{p}{1-p}}}$, following the same reasoning used in the weight assignments of Eq.~\eqref{optimalRBIM}.

\subsection{Probability of error-correction string given 3D syndromes}
\indent In principle, an optimal decoder can be constructed by considering all possible error strings $E$ consistent with a given set of measured anyon syndromes in 3D. Similar to Eq.~\eqref{Bayes} in the main text, the probability of the homology classes can be calculated by
\begin{equation}
    P(h|{\bm{s}}) \propto \sum_{E\in h} P(E|{\bm{s}}) = \sum_{E\in h} \frac{P(\bm{s}|E)P(E)}{P(\bm{s})} \propto \sum_{E\in h} P(\bm{s}|E)P(E),
    \label{measurement_optimal}
\end{equation}
where $\bm{s}$ denotes all terminals for the error-correction strings. The probability of each string $E$ is given by
\begin{equation}
    P(E)=\left(\frac{p}{1-p}\right)^{|E_{space}|}\left(\frac{q}{1-q}\right)^{|E_{time}|},
    \label{measurement_PE}
\end{equation}
where $p$ is the pair-creation error rate, $q$ is the measurement error rate, and $|E_{space}|$ and $|E_{time}|$ denote the lengths of the spacelike and timelike components of the error string, respectively. Obviously, $|E|=|E_{space}|+|E_{time}|$.\\
\indent Accounting for multiple rounds of physical error and measurements from time $t_i$ to $t_f$, and analogous to Eq.~\eqref{Probability}, $P(\bm{s}|E)$ can be calculated as the norm of the state
\begin{align}
    &\prod_{l} \left[ (1-\lambda_l)(1-A_l) + \lambda_lA_l \right]_{t_f} \hat{E}_{t_f} \times \cdots \times\\
    &\prod_{k} \left[ (1-\lambda_k)(1-A_k) + \lambda_kA_k \right]_{t_i+1} \hat{E}_{t_i+1}\ket{\psi} \prod_{j} \left[ (1-\lambda_j)(1-A_j) + \lambda_jA_j \right]_{t_i} \hat{E}_{t_i}\ket{\psi}\\
    =& \prod_{t=t_i}^{t_f} \mathcal{T} \{ \prod_{j} \left[ (1-\lambda_j)(1-A_j) + \lambda_jA_j \right]_{t} \hat{E}_{t}\}\ket{\psi},
    \label{measurement_Probability}
\end{align}
where $\ket{\psi}$ is the state in which quantum information is encoded, $A_{j,k,l}$ are commuting projectors that define the TO as in Eq.~\eqref{commuting_projector}, $\hat{E}_t$ is the physical error according to error-correction string $E$ at time $t$, and $\mathcal{T} \{\ldots\}$ denotes the time-ordered product.
\end{document}